\documentclass[10pt,journal,compsoc]{IEEEtran}
\usepackage{amsmath,amsfonts}

\usepackage[ruled,linesnumbered,algo2e]{algorithm2e} 
\usepackage{algorithmic}
\usepackage{algorithm}  
\usepackage{array}
\usepackage{subfig} 
\usepackage{textcomp}
\usepackage{url}
\usepackage{verbatim}
\usepackage{graphicx}
\usepackage{cite}
\usepackage{mathrsfs}
\usepackage{booktabs}

\usepackage{lineno}
\usepackage{multirow}
\usepackage{amsthm}

\usepackage{color}
\usepackage{ragged2e}
\renewcommand{\justify}{\leftskip=0pt \rightskip=0pt plus 0cm}

\usepackage[inkscapelatex=false]{svg}
\usepackage{threeparttable}
\usepackage{diagbox}
\usepackage[normalem]{ulem}

\usepackage{soul,xcolor}
\setstcolor{red}

\newtheorem{definition}{Definition}
\newtheorem{problem}{Problem}

\begin{document}

\title{Attention Is Not the Only Choice: Counterfactual Reasoning for Path-Based Explainable Recommendation}

\author{\IEEEauthorblockN{Yicong Li\IEEEauthorrefmark{1},
Xiangguo Sun\IEEEauthorrefmark{1} \IEEEauthorrefmark{2},
Hongxu Chen, 
Sixiao Zhang, 
Yu Yang, 
Guandong Xu \IEEEauthorrefmark{2}, \IEEEmembership{Member, IEEE} }

\IEEEcompsocitemizethanks{\IEEEcompsocthanksitem Yicong Li and Guandong Xu are with Data Science and Machine Intelligence Lab, Faculty of Engineering and Information Technology, University of Technology Sydney, Sydney, New South Wales, 2007, Australia. E-mail: Yicong.Li@student.uts.edu.au, Guandong.Xu@uts.edu.au
\IEEEcompsocthanksitem Xiangguo Sun is with Department of Systems Engineering and Engineering Management, The Chinese University of Hong Kong, Hong Kong SAR, 999077, China. E-mail: xgsun@se.cuhk.edu.hk
\IEEEcompsocthanksitem Hongxu Chen is with School of Information Technology and Electrical Engineering, The University of Queensland, Brisbane, Queensland, 4072, Australia. E-mail: Hongxu.Chen@uq.edu.au
\IEEEcompsocthanksitem Sixiao Zhang is with Department of Systems Engineering and Engineering Management, Nanyang Technological University, Singapore, 999077, Singapore. E-mail: sixiao001@e.ntu.edu.sg 
\IEEEcompsocthanksitem Yu Yang is with Department of Computing, The Hong Kong Polytechnic University, Hong Kong SAR, 999077, China. E-mail: cs-yu.yang@polyu.edu.hk 
}
\thanks{* Both authors contributed equally to this research.}
\thanks{\dag Corresponding author.}}

\markboth{IEEE TRANSACTIONS ON KNOWLEDGE AND DATA ENGINEERING}%
{IEEE TRANSACTIONS ON KNOWLEDGE AND DATA ENGINEERING}

\IEEEtitleabstractindextext{%
\begin{abstract}
\justify{
Compared with only pursuing recommendation accuracy, the explainability of a recommendation model has drawn more attention in recent years. Many graph-based recommendations resort to informative paths with the attention mechanism for the explanation. 
Unfortunately, these attention weights are intentionally designed for model accuracy but not explainability. 
Recently, some researchers have started to question attention-based explainability because the attention weights are unstable for different reproductions, and they may not always align with human intuition.
Inspired by the counterfactual reasoning from causality learning theory, we propose a novel explainable framework targeting path-based recommendations, wherein the explainable weights of paths are learned to replace attention weights. Specifically, we design two counterfactual reasoning algorithms from both path representation and path topological structure perspectives. Moreover, unlike traditional case studies, we also propose a package of explainability evaluation solutions with both qualitative and quantitative methods. We conduct extensive experiments on four real-world datasets, the results of which further demonstrate the effectiveness and reliability of our method. 
}
\end{abstract}

\begin{IEEEkeywords}
Counterfactual Reasoning, Path-based Recommendation, Explainable Recommendation.
\end{IEEEkeywords}}

\maketitle

\section{Introduction}

Nowadays, recommendation systems driven by knowledge graphs have achieved promising performance through various advanced graph neural networks \cite{li2021hyperbolic, hu2018leveraging,sun2023all}. 
However, the complexity of graph structure and rich information also brings a huge challenge to the model interpretability \cite{zhang2020explainable} compared with traditional recommendations such as collaborative filtering-based models.

To solve this problem, many researchers leverage graph-based models for recommendation and then explain their results via significant paths to the target items because they believe these paths can preserve real-world causality. Unfortunately, there are usually a lot of underlying paths to impact the final decision, which brings a huge challenge to select more explainable paths from the large candidate space. Recently, some work \cite{wang2019kgat, chen2021temporal, ma2022kr, li2023reinforcement, zhao2022time} have integrated the attention mechanisms into their models and learned the path weights for further explanation. These weights are usually reflected in an item's one-hop neighbors \cite{wang2019kgat}, users' purchase records \cite{chen2021temporal, li2023reinforcement}, and some external knowledge \cite{ma2022kr, zhao2022time}, and can be evaluated by some visualization cases.

\begin{figure}[ht]
	\centering
	\includegraphics[width=0.45\textwidth]{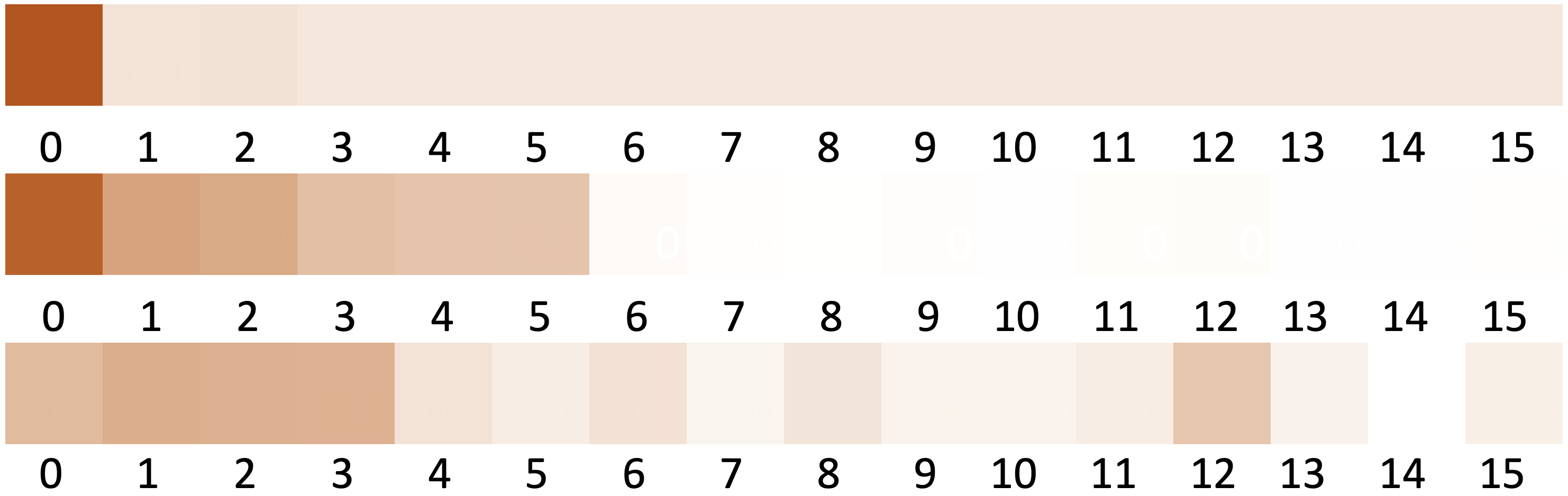}
	\caption{The attention weights on 16 paths via three times independent training. Each block presents a path, and darker orange means higher weight.} \label{fig:intro_att_weight}
\end{figure}

However, the concerns are mainly concluded as the following two aspects: 
\textbf{{Firstly}}, the attention mechanism is widely questioned for its interpretability because many studies have found the weak reliability that the attention mechanism performs \cite{wiegreffe2019attention, serrano2019attention, brunner2019identifiability, grimsley2020attention}. As shown in Figure \ref{fig:intro_att_weight}, we run an attention-based model on 16 underlying paths three independent times and draw the path-level attention weights in the heat map where each block denotes a specific path and the darker orange color means higher attention weight. We can see that the attention-based model can not ensure stable weight distributions via different independent running, which is far-fetched to convince customers to accept the explanation for the recommendation. 
\textbf{{Secondly}}, the attention mechanism used in graphs also tends to assign higher weights to those common frequent paths, which usually carry quite general and wide information. 
Whereas, those particular paths that carry a significant amount of explanation-specific semantics are not well captured (see further discussion on this phenomenon in section \ref{subsec:casestudy}). 

A promising direction to overcome the above problems is counterfactual reasoning. Generally speaking, counterfactual reasoning follows the ``what-if'' thinking: what will happen to the result if a condition does not hold anymore \cite{molnar2020interpretable}. In particular, a condition that causes the final result hugely changes is regarded as an important reason. Inspired by this, in the recommendation scenario, we can try to find causes by adding some slight disturbance to the candidate clues to see which clue perturbation makes the recommendation score change. If a slight disturbance of the path leads to a large decrease in the recommendation score, the current path is regarded as a significant one, and a small decrease indicates a less important one.
This new direction might benefit us with the following potentials: \textbf{First of all}, counterfactual reasoning mainly cares about model input and output, and goes beyond the inside mechanism differences among different models, which is perfect to support a model-agnostic explanation. This means we can use the same way on multiple recommendation models to see which one is more trustworthy; 
\textbf{Secondly}, the counterfactual mechanism can be more effective for those informative clues because general (wide) clues usually have lower uncertainty and they are more difficult to be disturbed to flip the recommendation score. This is promising to overcome the shortcoming of attention-based explanations that may easily overlook those informative paths.

Recent researches \cite{tan2021counterfactual, kaffes2021model, xiong2021counterfactual, ranjbar2023explaining} have started to explore the feasibility of counterfactual reasoning in recommendation explanation. However, most of them focus on the item \cite{kaffes2021model} alone, the item's aspect feature \cite{tan2021counterfactual, ranjbar2023explaining}, or user feature \cite{xiong2021counterfactual}. They are not designed for the path-based explanation, which is one of the most convincing evidence forms for graph-structural data. Although \cite{ghazimatin2020prince} explores the counterfactual reasoning on the knowledge graph, the reasoning weights are tightly tailored to the proposed recommendation, which cannot be used to evaluate different recommendation backends.

To fill the gap between counterfactual reasoning and the explanation of path-based recommendation models, we propose a Counterfactual Path-based Explainable Recommendation (CPER for short). We design two effective counterfactual reasoning methods to find explainable paths from the perspective of their representations and the topological structures. In particular, we propose an optimization framework to learn appropriate perturbation factors on path embeddings (section \ref{subsubsec:high-dim_count}). Meanwhile, we conduct counterfactual reasoning on path topological structure by learning a path manipulation policy driven by reinforcement learning (section \ref{subsubsec:rl_count}).  
Besides, unlike traditional work that mostly evaluates their explanation via case studies, we propose a package of solutions to evaluate the explainability with both \textit{qualitative} and \textit{quantitative} ways (section \ref{subsec:eva_explain}). 
We evaluate our explanation framework on four real-world datasets with a very typical path-based recommendation backend, which reveals the significant advantages of our method. In summary, our main contributions are as follows:

\begin{itemize}
    \item We propose a novel explainable framework for path-based recommendation via counterfactual reasoning conducted on both path representation and path topological structure.
    \item For the path topological structure-based counterfactual reasoning, we novelly devise a reinforcement learning method to learn a path manipulation strategy for perturbation in counterfactual learning.
    \item We propose a package of solutions to evaluate explainability quality for path-based recommendations. Unlike traditional attention-based explanation, which is mostly evaluated by case studies, our proposed measurements include both \textbf{\textit{qualitative}} and \textbf{\textit{quantitative}} methods, which can be widely used in comparing various explanation methods.  
    \item Extensive experiments conducted on four real-world datasets further demonstrate the effectiveness of our framework. We carefully compare our explanations with attention-based ones and the results reveal that our explanation method has higher stability, effectiveness, and confidence.
\end{itemize}

\section{Preliminaries}

\subsection{Problem Definition}
\label{sec:problem_def}

\begin{definition}
\textbf{Recommendation Graph.} Recommendation graph $G=(V,E)$ is a heterogeneous graph that contains three types of vertices including users $U=\{u_1, u_2, ...,u_m\}$, items $I=\{i_1, i_2, ..., i_n\}$ and items' attributes $A=\{a_1, a_2,..., a_q\}$, where $V=(U,I,A)$. $e\in E$ is an edge connecting any pair of vertices in $V$. 
\end{definition}

\begin{definition}
    \textbf{Path on Recommendation Graph.} Path $P_{u_m \rightarrow i_n}$ is all connected vertex sequences starting at a user $u_m$ and ending at an item $i_n$, indicating the relationship among users, items, and item attributes.
\end{definition}

Given the recommendation graph $G$, we generate the paths $P$ according to Algorithm \ref{algorithm:t-randomwalk}. Specifically, for each user and item vertex, we explore the paths according to RandomWalk \cite{xia2019random}. Furthermore, explainable paths are defined as paths with 4 to 6 lengths from users/items to items. 
For example, one of the explainable paths $p$ might be $u_1 \rightarrow i_1 \rightarrow a_1 \rightarrow i_2 \rightarrow a_2 \rightarrow i_3 $, which is regarded as the explanation of purchasing item $i_3$.

\begin{problem}
\textbf{(Counterfactual Path-based Explainable Recommendation)} Given a recommendation graph $G=(V,E)$, the path-based explainable recommendation aims to predict each user $u_m$'s preferred $l$ items $I^{u_m} = \{i^{u_m}_1, i^{u_m}_2,..., i^{u_m}_l\}\in I$ and simultaneously 
generate paths $P_{\rightarrow I^{u_m}}$ where the end nodes are the predicted items $I^{u_m}$. Counterfactual-based explainable recommendation leverages counterfactual ideas for a further selection of the counterfactual paths $\mathcal{C}_{\rightarrow I^{u_m}}$ for the assistance of paths explainability enhancement. Therefore, the objective of counterfactual-based explainable recommendation is to provide recommendations for $I^{u_m}$ and learn the underlying counterfactual paths $\mathcal{C}_{\rightarrow I^{u_m}}$ for a more convincing explanation. 
\end{problem}

For the above example explainable path, when performing counterfactual learning on path $p$, it would be slightly perturbed and change to $p^{\prime}$, like $u_1 \rightarrow i_{100} \rightarrow a_1 \rightarrow i_2 \rightarrow a_2 \rightarrow i_3 $, if this counterfactual path $p^{\prime}$ significantly influences the recommendation prediction, it is regarded as an explainable path with high importance.

\subsection{Why Counterfactual Reasoning Works?}
Intuitively, most paths in the recommendation graph are very general, and only a few of them are inspiring for the explanation. For example, in the \textit{Amazon Musical Instrument} dataset, the category \textit{Musical Instrument} obviously holds the widest information and has the most link throughout all the category nodes. This leads to the paths related to the \textit{Musical Instrument} category the most. However, 
from the perspective of information theory \cite{brillouin2013science}, those general paths appearing very frequently usually contain a very limited amount of information (a.k.a \textbf{\textit{lower uncertainty}}), which are less helpful for the more informative explanation. Traditional attention mechanism intends to assign higher weights to those highly frequent paths but ignores those more informative paths with {\textit{higher uncertainty}}. Unlike traditional attention mechanisms, counterfactual reasoning aims to add slight perturbation to these paths and see the consequence from the recommendation results \cite{tan2021counterfactual, kaffes2021model, ghazimatin2020prince}. Obviously, the general paths with {\textit{lower uncertainty}} will be more difficult to be affected, but those informative paths with {\textit{higher uncertainty}} can be easily disturbed, causing the results dramatically changed. Therefore, we can leverage this new approach to find more informative paths for recommendation explanation.

\begin{algorithm}[H]
\caption{Paths Exploration on Recommendation Graph}\label{algorithm:t-randomwalk} 
\begin{algorithmic}[1]
\renewcommand{\algorithmicrequire}{\textbf{Input:}}
 \renewcommand{\algorithmicensure}{\textbf{Output:}}
 \REQUIRE user set $U=\{u_1, u_2, ..., u_m\}$, item set $I=\{i_1,i_2,...,i_n\}$, item attribute set $A=\{a_1, a_2, ...,a_q\}, max\_path\_length, min\_path\_length$
 \ENSURE  path set $P$

 $P \leftarrow new\ List$\;
 
 \For{vertex $v \in \{U,I\}$}
    { $path\leftarrow new\ List$\;
    
     $path.append(v)$\;
          
     \While{$l<max\_path\_length$ }{
        
        $v^{\prime}\leftarrow Random(Neighbor(path[-1]))$\;
        
        $edge\leftarrow [path[-1],v^{\prime}]$\;
        
        $path.append(v^{\prime})$\;

        $l\leftarrow len(path)\;$

        \If{$l > min\_path\_length$ and $path[-1]\in Neighbor(v)$}{
        $P.append(path)\;$
    }
    }
}
\end{algorithmic}
\end{algorithm}

 

    
     
     
        
        

\section{Counterfactual Reasoning on Paths}
\label{subsec:count_paths}
To avoid the unreliability issue brought by attention-based explainable weights of paths, we learn the explainable path weights via counterfactual learning to replace the traditional attention weights. The main idea is that once performing perturbation $\gamma$ on each explainable path $p$ or the path set $P$, the decreasing scores of the recommendation prediction score can be regarded as the explainable weight of the path or the path set. 
To achieve this, we conduct counterfactual reasoning from two perturbation perspectives, including path representation and path topological structure. 
Specifically, we conduct counterfactual reasoning on path representations by learning appropriate perturbation factors on path embeddings, and counterfactual reasoning on path topological structure via learning a reinforcement learning-based strategy to manipulate paths by replacing some path vertices.

\subsection{Counterfactual Reasoning on Path Representations}
\label{subsubsec:high-dim_count}

Let $x$ be the path representation and $x^{\prime}$ be a disturbed representation generated by counterfactual reasoning on the same path. If we feed this new representation $x^{\prime}$ to the recommendation backend model $R(h_U, h_I, \{x_1,x_2,...,x_p\}, \Theta)$, it can output a new recommendation prediction score $s^{\prime}$ compared with the original $s$. Mathematically,
\begin{gather} \label{eq:s_R}
    s = R(h_U, h_I, \{x_1,x_2,...,x_p\}, \Theta) \\
    s^{\prime} = R(h_U, h_I, \{x_1^{\prime},x_2,...,x_p\}, \Theta)
\end{gather}
where path representation $x_1$ is disturbed to $x^{\prime}$, and $R$ is introduced in section \ref{subsec:rec}, which can be calculated according to Eq. \ref{eq:new_item_update}, \ref{eq:seq_pref}, \ref{eq:cal_s}.

Unlike traditional attention-based explanation, which measures the path importance by attention weights, we evaluate the path significance from two aspects: how much is this perturbation added to the input $dis(x, x^{\prime})$, and how much has the result changed $dis(s, s^{\prime})$. If the perturbation on a path is very slight but causes a dramatic decrease in results, the corresponding path should be very informative and important. To this end, this section aims to learn a slight perturbation factor and at the same time find informative paths affected by this perturbation.

\begin{figure}[h]
	\centering \includegraphics[width=0.45\textwidth]{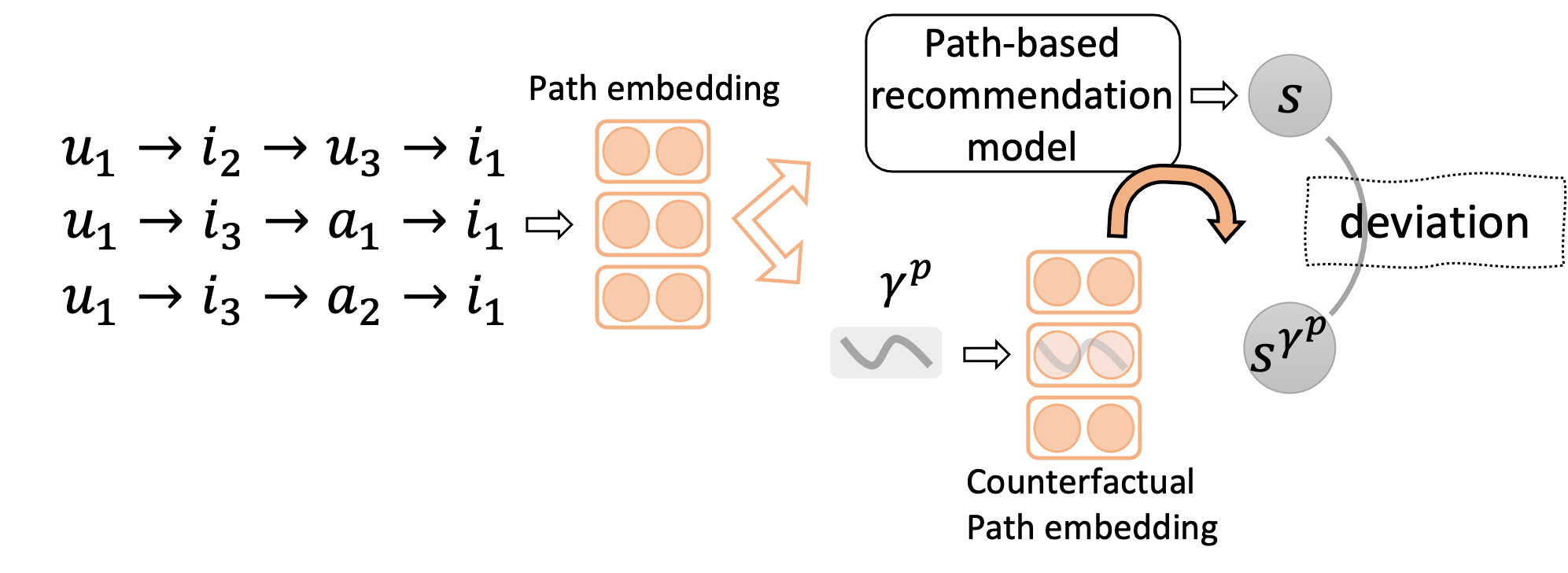}
	\caption{Counterfactual reasoning on path representations.}\label{fig:counterfactual_path_high_dim}
\end{figure}

To realize the above two targets, we design slight perturbations on the path embeddings as shown in Figure \ref{fig:counterfactual_path_high_dim}. Specifically, the perturbation $\gamma ^p$ is added to the path embeddings to get the counterfactual path embeddings. After feeding them to the path-based recommendation backend, the prediction score would change to $s^{\gamma ^p}$ from the original score $s$. The deviation is regarded as the explainable weight in our work. For learning the slight perturbation, we minimize $dis(x, x^{\prime})$ as follows:
\begin{equation}
    \ell_1 = || \gamma ^{p} ||_2^2 + 
    \alpha || \gamma ^{p} ||_1
\end{equation}
where $\gamma ^{p}$ is the perturbation vector on the path $p$. $||.||_2$ and $||.||_1$ are the L2 norm and L1 norm, respectively. $\alpha$ is the hyperparameter to balance the L2 norm and L1 norm. 
The above loss $\ell_1$ is to minimize the perturbation on the path embedding of path $p$. Besides, another target is to maximize the influence $dis(s, s^{\prime})$ after perturbing the original paths, which can be ensured by minimizing the following target:
\begin{equation}
    \ell_2 = -s_{u,i} + s_{u,i}^{\gamma ^{p}}
\end{equation}
where $s_{u, i}$ is the original recommendation score for user $u$ and item $i$;  $s_{u,i}^{\gamma ^{p}}$ is the new output of the recommendation model after we change the representation of path $p$. 
Considering $\ell_2$ is not differentiable when negative, we adopt a hinge loss to relax the constraint. 
Then the final loss from user $u$ to item $i$ is calculated as follows:
\begin{equation}\label{equ:final_loss}
    \mathcal{L}_{u,i} = \ell_1 + \lambda max(0, \beta + \ell_2) 
\end{equation}
where $\lambda$ and $\beta$ are the hyperparameters to trade off the weight of each term. Further, for learning the whole explainable model, the objective is to learn slight perturbations as follows. 
\begin{equation}\label{eq:optimization_perturb}
    \Gamma^{\star}=\underset{\Gamma}{\operatorname{argmin}} \left(\mathbb{E}_{u \in \mathcal{U}, i \in \mathcal{I}} \left[\mathcal{L}_{u,i}\right]\right)
\end{equation}
where $\Gamma^{\star}=[\gamma_{u_1,i_1}^{p_1}, \cdots, \gamma_{u_m,i_n}^{p_k}]$ contains the optimal perturbations for each path related to user-item pairs trained according to the loss in equation \ref{equ:final_loss}. Then, perturbations are performed on each path respectively, and we choose the paths that have an impaired influence on the recommendation score ($s_{u,i}-s_{u,i}^{\gamma ^p} >0$) as explainable paths. This score also indicates the impact and explainable weight of the path quantitatively.

\subsection{Counterfactual Reasoning on Path Topological Structure}

\label{subsubsec:rl_count}
In addition to the high-dimensional perturbation approach, we also develop another method to perturb topological path structures. To achieve this goal, we aim to enhance the interpretability of the perturbation, as compared to perturbation path representations. Specifically, we devise a path manipulation strategy using reinforcement learning, taking advantage of its powerful search capabilities.


\begin{figure}[h]
	\centering
	\includegraphics[width=0.45\textwidth]{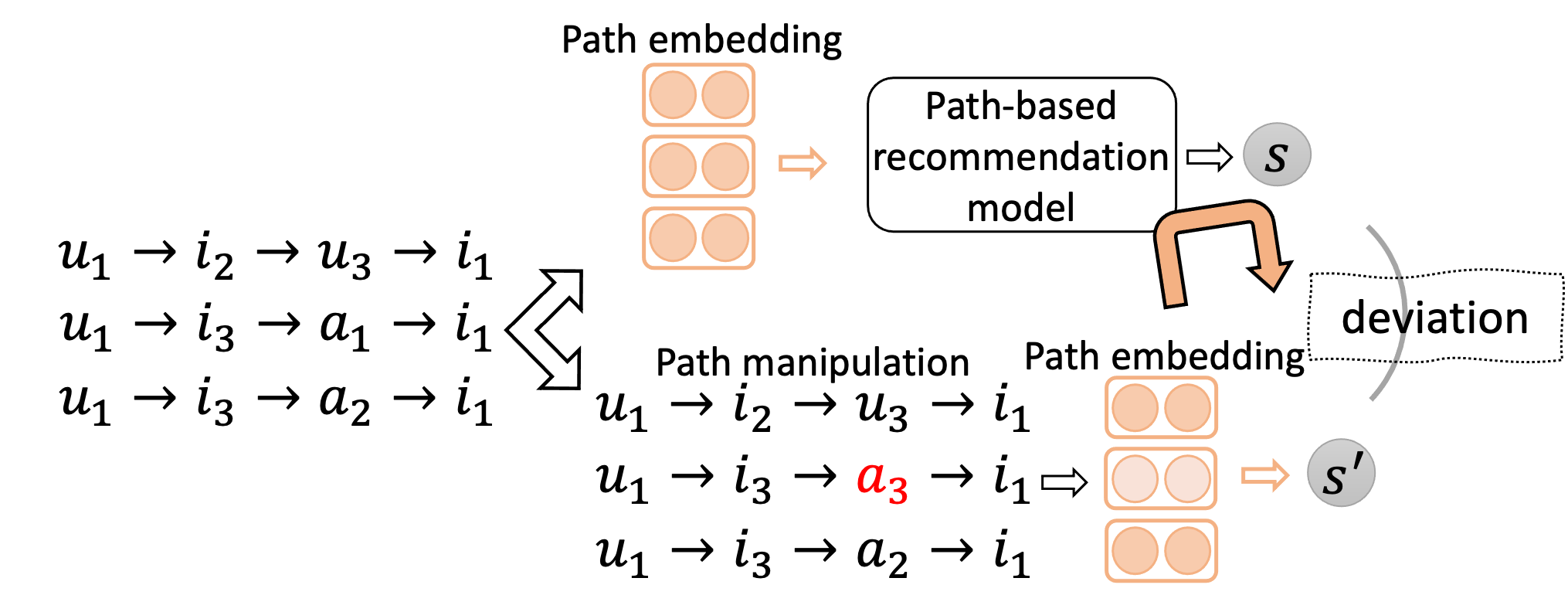}
	\caption{Counterfactual reasoning on path structure. }  
	\label{fig:counterfactual_path_rl} 
\end{figure}

\textbf{Reinforcement Learning Framework.}
Let us consider a path set that will be fed into a path-based recommendation model. We can try to replace some vertices on paths and see what will happen to the model result. We hope the agent finds a path-level operation sequence with fewer manipulations but makes the model's output score decrease as much as possible. This operation sequence can be treated as a trajectory. The agent's current action is dependent on its previously changed path set and the model output score, which can be approximated to a Markov Decision Process (MDP): $(\mathcal{S}, \mathcal{A}, \mathcal{T}, \mathcal{W})$ where $\mathcal{S}$ is the state space, $\mathcal{A}$ is the action space, $\mathcal{T}$ is the state transition pattern,  and $\mathcal{W}$ is the reward. $\mathcal{O}$ is the optimization of the model. 
In this paper, the recommendation system is the environment; each time the agent manipulates the path set and feeds them into the recommendation model, leading to a new path set state and a new predicted score. Therefore the path set state and the predicted score together serve as the environment state. The state transition $\mathcal{T}$ depends on the recommendation system.

\textbf{States.}
The state $\mathcal{s} \in \mathcal{S}$ is defined as a combination of path set (collection) and the model output score like $(P_{\rightarrow i_n^{u}}, s_{u,i})$, where $P_{\rightarrow i_n^{u}}$ denotes all the paths to the item $i_n$ for user $u$. Taking Figure \ref{fig:counterfactual_path_rl} as an example, the first path to item $i_1$ for user $u_1$ can be denoted as $p^1_{\rightarrow r_1^{u_1}}=(u_1, i_2, u_3, i_1)$. Here, the initial state $\mathcal{s}_0$ includes the tuple of all the paths and the original model score for the user-item pair $(u, i_n)$. Each time when the agent manipulates the path set collection, the modified collection together with the corresponding model score will be treated as a new state.

\textbf{Actions.}
For a path collection $P=\{p_1,p_2,\cdots, p_K\}$, the agent tries to select several vertex positions from $P$ and replace them with some alternative vertex in $\mathcal{V}$.
Therefore, the action space can be denoted by $\mathcal{A}=P\times \mathcal{V}$. We also set the constraint that the perturbed vertex should keep the same node type as the original one to keep the perturbation smallest to the most extent. 
Further, to reduce the searching space, we let the alternative choices of each vertex on the path be its two-hop neighboring nodes with the same node type in the knowledge graph. The rationale for randomly selecting two-hop neighboring nodes as the candidate set is that two-hop neighboring vertex tends to have similar information to the original vertex.
For example, we can define $\mathcal{N}=\{\mathcal{N}_1,\mathcal{N}_2,\cdots, \mathcal{N}_\mathcal{V}\}$ as the alternative set for the agent where $\mathcal{N}_v \in \mathcal{N}$ contains node $v$'s two-hop neighboring nodes. 
Since most graph data follow the power-law distribution, the average size of $\mathcal{N}_v$ can be very small like $k\ll |\mathcal{V}|$, and then the action space can be reduced from $|\mathcal{P}| \times |\mathcal{V}|$ to $k|\mathcal{P}|$. Specifically, whether to select a vertex to be replaced is decided by a designed neural network, and the replacing vertex process from $\mathcal{N}$ is a random selection.

\textbf{Transition.} The transition $\mathcal{T}$ presents the transition probability from the current state to the next state, which can be denoted as a map function $\mathcal{T}: \mathcal{S} \times \mathcal{A} \times  \mathcal{S} \rightarrow [0,1]$.  

\textbf{Rewards.} To calculate the reward, we first feed the paths into the fixed recommendation model to predict the positive user-item score $s_{u,i}$. Then, the reward can be calculated as follows.
    \begin{equation}
        \mathcal{W} = \left\{
        \begin{array}{lr}
            -\zeta \cdot |C| -\epsilon \cdot |F| + \eta \cdot \Delta s &, \Delta s > 0, |F| \neq 0 \\
            0 &, \Delta s < 0 \\
            -200 &,  |F| = 0 
        \end{array}
        \right.
    \label{eq:rl_reward}
    \end{equation}
where $|C|$ is the number of selected paths of the agent, and $|F|$ is the number of the replaced vertices. These two terms make sure the least perturbation of the path collection. $\zeta$, $\epsilon$ and $\eta$ are the positive coefficients. $\Delta s = s_{u,i} - s_{u,i}^c$, where $s_{u,i}$ and $s_{u,i}^c$ are the original user-item score and the affected user-item score by counterfactual paths, both of which can be obtained from the fixed recommendation backend. The rationale is to perturb the least nodes $|F|$ and paths $|C|$ and leads to the largest recommendation score drop.

\textbf{Optimization.} To learn optimal manipulation strategy, we follow \cite{williams1992simple} to train our agent model by maximizing the accumulated rewards:
\begin{equation} \label{eq:rl_loss}
        J(\theta) = \mathbb{E}_{\pi}(\mathcal{W})
	\end{equation}
where $\mathbb{E}$ is the expectation of the rewards. The least perturbation and also more difference $\Delta s$ means the more explainable weights for the original paths. Therefore, we choose the path set with the largest reward as the explainable path set.
For explainability evaluation, we combine the paths found by counterfactual reasoning on path representations and structures together and compare them with attention-based path explanation in subsection \ref{subsec:exp_quali_explainability} and \ref{subsec:exp_quan_explainability}. We further discuss these two proposed counterfactual reasoning methods in subsection \ref{sec:quantitative_analy_comp}. Specifically, after learning the explainable paths via counterfactual reasoning on path representations and on path topological structure, we will regard the intersection as the final explainable paths for evaluation in section \ref{subsubsec:confidence} and \ref{subsec:fedelity}.

\subsection{Evaluating the Explainability for Path-based Recommendation}\label{subsec:eva_explain}
Evaluating whether the explanation is trustworthy is very subjective. To the best of our knowledge, there is seldom a widely accepted measurement to evaluate path-based explainability. In this paper, we wish to push forward this area by proposing both qualitative and quantitative methods. 

\textbf{Qualitative Evaluation.} We evaluate the \uline{\textbf{\textit{{stability}}}} of the explainable method by independently repeating the recommendation model learning multiple times and seeing whether the explainable path distribution is consistent. The more stable the explainable distribution is, the more reliable of the explainable method is. (see in section \ref{subsubsection:stability}).
To evaluate the \uline{\textbf{\textit{{effectiveness}}}} of our explainable method, we randomly add an irrelevant path to the path set to see what weight the explanation framework learns (see in section \ref{subsubsection:effectiveness}). Intuitively, the irrelevant path's explainable weight should be as small as possible to make sense. Otherwise, the explainable weight is unreliable.

\textbf{Quantitative Evaluation.} We evaluate the \uline{\textbf{\textit{{confidence}}}} of explainability inspired by information theory, where high entropy means low certainty of the information. Therefore, we define the confidence of each explainable path as its uncertainty (a.k.a entropy), which can be calculated as follows: $-\sum_{k} c(p_k)logc(p_k)$, where $k$ is the number of paths, $c(p_k)$ is the weight of $k$-th path learned by attention mechanism or counterfactual method. Intuitively, a better explanation model tends to explore explainable paths more confidently and keeps the path uncertainty relatively lower. We also propose to evaluate the explainability via \uline{\textbf{\textit{{informativeness}}}}. Here we feed the learned explainable paths back to the recommendation backend to see how much the learned explainable paths contribute to the recommendation performance. The closer to the original result, the more informative the learned explainable paths are, compared with all the rest paths. We also use a frequently used metric, \uline{\textbf{\textit{{fidelity}}}}, to evaluate the explainability. It measures the decrease in the prediction score when removing different ratios of explainable paths from the input explainable paths. A larger fidelity value indicates stronger counterfactual weights and better explainability.

\section{Path-based Recommendation Backend}
\label{subsec:rec}

Here we design a lightweight path-based recommendation model as the backend. Note that this model can be seamlessly replaced by any other path-based recommendation model as long as they take paths as partial/total input.

\begin{figure}[h]
	\centering
	\includegraphics[width=0.45\textwidth]{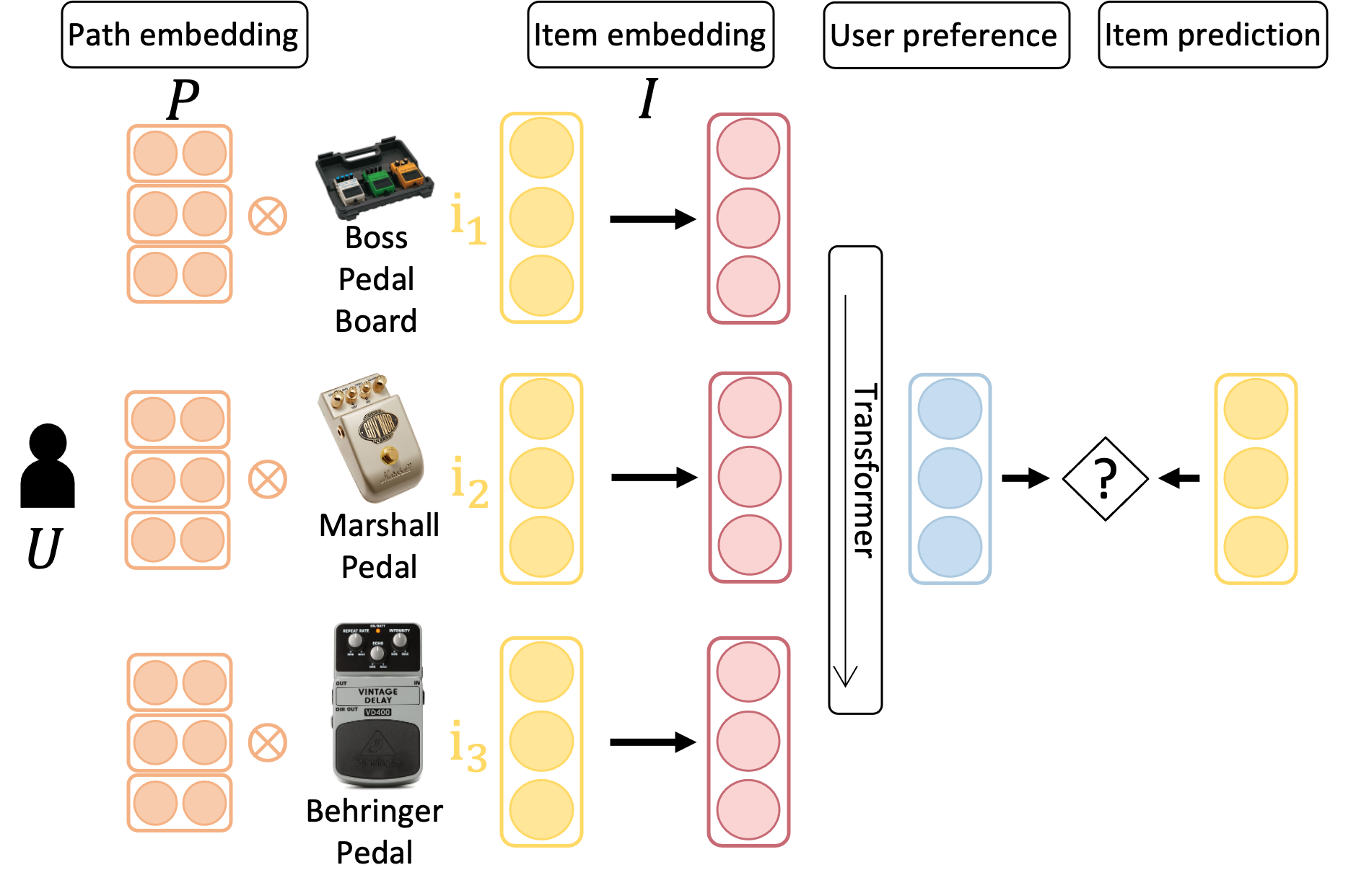}
	\caption{The sequential recommendation module. $\otimes$ represents two-layer attention for item embedding enhancement via related path embeddings.
	} \label{fig:rec} 
\end{figure}

The workflow of the recommendation backend is to model the users' behavior evolution and predict the user's preferred items. As shown in Figure \ref{fig:rec}, the predicted item scores $S$ can be obtained like $\mathcal{S}=(S|U,I,P,\Theta)$, which only relies on the users $U$, items $I$, paths $P$ and recommendation model parameters $\Theta$.
The initial user and item embeddings could be obtained via temporal DeepWalk, which is a variation of a classical graph representation learning algorithm, DeepWalk \cite{perozzi2014deepwalk}. In the traditional DeepWalk, it firstly explores paths from each node on the graph (RandomWalk) and then performs Skip-gram \cite{mikolov2013efficient} to obtain the nodes embeddings. However, DeepWalk only considers the static graph presentation learning due to the RandomWalk algorithm performed on the static graph, which overlooks the dynamic evolution of user-item purchase relations in recommendation data. Specifically, in the real sequential scenario \cite{chen2022intent, qiu2022contrastive, fan2022sequential}, the future action (user's purchase behavior) is unknown, but in the static recommendation graph, it ignores the behavior's timestamp and regards the future behavior as the known action. This makes the presentation learning disobey real situations. Therefore, we propose a temporal RandomWalk to consider the temporal user-item purchase behavior evolution to generate temporal paths to ensure that there are no future behaviors along the path. To realize the purpose, we guarantee that the latter edges' (relations) timestamps are less than that of the former edges (relations) along each path. Then, we also adopt Skip-gram in the following step to learn the nodes embeddings. For the path embeddings, we use an average pooling on the related node embeddings to obtain. 
And we generate paths via the random path exploration for each item with the same settings in \cite{chen2021temporal}. Specifically, assuming that user $u_m$ has purchased items including $\{i_1^{u_m}, i_2^{u_m}, ..., i_o^{u_m}\}$, for each item $i^{u_m}$, we extract paths starting from user $u_m$ and ending at item $i^{u_m}$. 
The path embeddings can be easily obtained via the average pooling on the path node embeddings. Then the item embeddings can be updated by its representation and its associated paths as follows:
\begin{equation}\label{eq:new_item_update}
    h_{i_o^{u_m}}^{\prime} = f(W_{i_o^{u_m}} h_{i_o^{u_m}} + \sum_{z=1}^{Z} W_{p_z} h_{p_z} +b)\odot h_{i_o^{u_m}}
\end{equation}
where $f(.)$ is ReLU activation function and $h_{i_o^{u_m}}$ is item $i_o^{u_m}$'s initial embedding. $W$ and $b$ are the weight matrix and bias vector, respectively. $p_z$ means the related $Z$ paths obtained in the last step, and $h_{i_o^{u_m}}^{\prime}$ is the updated item embedding. 
With the above formula, we can infer the user's preference as follows: 
\begin{equation}\label{eq:seq_pref}
    h_{u_m}^{seq}=g(h_{i_1^{u_m}}^{\prime}, h_{i_2^{u_m}}^{\prime}, ..., h_{i_o^{u_m}}^{\prime})
\end{equation}
where $h_{u_m}^{seq}$ represents user's preference; $g(.)$ is a sequential model such as Transformer \cite{vaswani2017attention}. Finally, the predicted score between user $u_m$ and item $i$ can be calculated with a Multilayer Perceptron (MLP) unit as follows:
\begin{equation}\label{eq:cal_s}
    s_{u_m,i} = MLP(h_{u_m}^{seq})
\end{equation}
According to \cite{hu2018leveraging, tang2015line}, the recommendation model can be effectively trained via implicit feedback loss with negative sampling: 
\begin{equation}
    \mathcal{L}_{u,i^+}=-log\ s_{u_m,i^+} -E_{i^- \sim D_{neg}}[log(1-s_{u,i^-})]
\end{equation}
where the first term models the positive user-item score and the second term models the negative feedback. Each positive pair $(u,i^+)$ is associated with 1 negative pair $(u,i^-)$, which can be sampled from a uniform noisy distribution $D_{neg}$. 

\section{Model Optimization}
\vspace{0.5em}

For the counterfactual learning on path representations in section \ref{subsubsec:high-dim_count}, it optimizes the loss to learn the best perturbation embedding for each user-item pair according to Eq. \ref{equ:final_loss}, and optimizes all perturbation embeddings for all user-item pairs according to Eq. \ref{eq:optimization_perturb}. 
For the counterfactual learning on path topological structure in section \ref{subsubsec:rl_count}, it optimizes the reward (Eq. \ref{eq:rl_reward}) of reinforcement learning by Eq. \ref{eq:rl_loss} to optimize the best explainable path set.

In the overall learning procedure, it should first train the black-box path-based recommendation backend $R$ in section \ref{subsec:rec}. Then, the two counterfactual reasoning algorithms in sections \ref{subsubsec:high-dim_count} and \ref{subsubsec:rl_count} are trained respectively. Specifically, for the counterfactual reasoning on path representations (section \ref{subsubsec:high-dim_count}), it performs the learned perturbation on each path representation and feeds them to the recommendation backend model $R$, and paths whose perturbed score are less than the original one are regarded as explainable paths. The deviation of the recommendation score is considered as the importance weight of each path. For the counterfactual reasoning on path topological structure (section \ref{subsubsec:rl_count}), it adopts reinforcement learning to manipulate vertexes on paths, and feed them to recommendation $R$. Likewise, it also uses the deviation score as the importance of paths.

\section{Experiment}
\label{sec:exp}


\subsection{Experimental Settings}
\subsubsection{\textbf{Datasets}}

We evaluate our recommendation method on four real-world datasets, named Amazon Musical Instruments, Amazon Automotive, Amazon Toys and Games, and MovieLens. The statistics are shown in Table \ref{tab:Dataset_Information}. They are subsets of a public product dataset, Amazon \cite{ni2019justifying}, which contains 29 categories of products from May 1996 to July 2014. This rich dataset also includes user-product interaction, product metadata and reviews from users. For better training, we only keep items bought more than 5 times and choose brand and category as item attributes. 
The MovieLens dataset is extracted from the public MovieLens 25M Dataset. 
For meta information alignment, we use the genre and tag as movies' meta information.

\subsubsection{\textbf{Baselines}}
We first compare our proposed CPER with the following 7 state-of-the-art baselines for the recommendation performance evaluation. Following are the details of them.
\begin{itemize}
    \item \textbf{CountER}. \cite{tan2021counterfactual} It adopts the counterfactual idea on items' aspects/attributes in recommendations and proposes a model-agnostic explainable recommendation with aspects-explanation. 
    \item \textbf{PGPR}. \cite{zhang2017joint, xian2019reinforcement} The work leverages reinforcement learning to explore paths and also predict preferred items, which is a model-specific explainable recommendation.
    \item \textbf{ACVAE}. \cite{xie2021adversarial} The method involves adversarial Variational Bayes with a contrastive loss for modeling the user interaction history in recommendation.
    \item \textbf{GRU4Rec}. \cite{hidasi2015session} It is a classical session recommendation, and the first work to use recurrent neural network (RNN) for user-item interaction history session modeling in recommender systems. 
    \item \textbf{Bert4Rec}. \cite{sun2019bert4rec} The work proposes to adopt bidirectional encoder representations from transformer in sequential recommender systems. 
    \item \textbf{SASRec}. \cite{kang2018self} It relies on self-attention mechanism to improve the user-item interaction history learning for sequential recommendations.
    \item \textbf{SRGNN}. \cite{wu2019session} The work uses graph neural networks with local interest and main purpose modeling for sequential recommendation enhancement.
\end{itemize}


Then for the explainability evaluation of CPER, we compare our explanation framework with attention-based explanations learned by the proposed path-based recommendation backend, qualitatively and quantitatively via evaluations introduced in subsection \ref{subsec:eva_explain}. 



\subsubsection{\textbf{Implement Details}}
Some sequential recommendation baselines, including GRU4Rec, Bert4Rec, SASRec and SRGNN, are implemented by the RecBole framework \cite{recbole1.0}. For the rest baselines, we use the code implemented by the authors.
All the comparison methods set the dimension as 100, keeping the same as CPER. For the Musical Instruments and Automotive datasets, the learning rate is 1e-3, for the Toys and Games dataset, the learning rate is 1e-4, and for the MovieLens dataset, the learning rate is 1e-5. 

In our baseline comparison experiment in subsection \ref{subsec:exp_overall_performance}, we adopt the default parameter settings from RecBole framework \cite{recbole1.0}, for GRU4Rec, Bert4Rec, SASRec and SRGNN. The rest baselines also keep the same parameter as the GitHub version implemented by the authors. 

As for testing the effectiveness of different components of our proposed recommendation backend in subsection \ref{subsec:exp_ablation_test}, we conduct the experiment on the Amazon Musical Instruments dataset, and the learning rate is 1e-3. The training epoch is set to 300 for all variants.

When testing the explainability of the overall proposed counterfactual reasoning in subsection  \ref{subsec:exp_quan_explainability} and \ref{subsec:exp_quali_explainability},
we conduct six experiments, including stability, effectiveness, confidence, informativeness, fidelity, and analysis of the reinforcement learning training process. The hyperparameters $\{\alpha$, $\beta$, $\lambda \}$ in counterfactual reasoning on path representations are $\{0.1, 0.5, 5\}$, respectively. For counterfactual reasoning on path topological structure, the hyperparameters $\{ \zeta$, $\epsilon$, $\eta \}$ are $\{10,10,100\}$, respectively. We train 50 epochs for both counterfactual reasoning models and combine the generated explainable paths for comparison with attention-based weights.

\begin{table}[ht]\centering
\resizebox{0.48\textwidth}{!}{
	\begin{tabular}{ccccccc}
		\hline \toprule
		\textbf{Datasets}& \textbf{User}& \textbf{Item} & \textbf{Category/Genre} & \textbf{Brand/Tag} \\ \hline
		Musical Instruments &1,450&11,457&429&1,185\\ 
		Automotive &4,600&36,663&1,592&3,790\\ 
		Toys and Games&9,300&58,743&820&5,404\\ 
            MovieLens&50,000&20,118&19&1,127&\\
        \bottomrule
	\end{tabular}}
	\caption{\label{tab:Dataset_Information}Dataset information.}
\end{table}

\subsubsection{\textbf{Research Questions}}
To evaluate the explainability and effectiveness of the proposed method, the experiment mainly answers the following questions. 

\begin{itemize}
    \item \textbf{RQ1}: From a \textit{quantitative} perspective, how is the explainability of the proposed counterfactual reasoning on paths compared with the traditional attention-based explainability?
    \item \textbf{RQ2}: From a \textit{qualitative} perspective, how is the explainability of the proposed counterfactual reasoning on paths compared with the traditional attention-based explainability?
    \item \textbf{RQ3}: How is the effectiveness of the proposed reinforcement learning-based path manipulation?
    \item \textbf{RQ4}: How does the proposed path-based recommendation perform compared with state-of-the-art backends, and how do the components contribute to the performance?
\end{itemize}

\subsection{Quantitative Evaluation for Explainability of our Counterfactual Reasoning (RQ1)}
\label{subsec:exp_quan_explainability}

In this section, we evaluate the explainability of our proposed counterfactual reasoning framework quantitatively. Specifically, we compare our method with traditional attention-based explanations using our previously introduced measurements in section \ref{subsec:eva_explain}: \textit{confidence}, \textit{informativeness} and \textit{fidelity}.

\subsubsection{\textbf{Confidence}}
\label{subsubsec:confidence}
Inspired by information theory \cite{brillouin2013science}, we introduce the uncertainty value (in subsection \ref{subsec:eva_explain}) to evaluate the explainability. 
Specifically, a higher uncertainty value means lower confidence in the generated explainable paths. 
Our learned explainable paths in this experiment are the intersection of two counterfactual learning methods for simplicity. 
To compare the confidence scores, we first randomly draw the generated path attention weights distribution, calculate the weight of the paths 3 times, and get the average entropy. Then we summarize multiple user-item pairs and present the overall uncertainty value in Table \ref{tab:certainty}. The entropy of attention weights probability and CPER paths probability are 2.25 and 1.87, respectively. It shows that our proposed method could learn lower entropy and more deterministic explanation results.

\begin{table}[ht]
\centering
\resizebox{0.4\textwidth}{!}{
\begin{tabular}{@{}lc@{}}
\hline \toprule
                      & Uncertainty Value \\ \hline
Attention-based Explanation & 2.25            \\
Our Explanation     & 1.87            \\ \bottomrule
\end{tabular}}
\caption{Uncertainty value.}
\vspace{-2em}
\label{tab:certainty}
\end{table}

\subsubsection{\textbf{Informativeness}}
\label{sec:quantitative_analy_comp}
We compare our two counterfactual reasoning methods with three baselines, including random selection, the attention module in section \ref{subsec:rec} and TMER \cite{chen2021temporal}, on the informativeness metric, which is defined in section \ref{subsec:eva_explain}.
Specifically, if an explainable path makes sense, the user's current purchasing action should be more related to this path rather than the others. Therefore, we only feed the learned explainable paths back to the recommendation backend, and check the new prediction score. If the new score is very close to the original score, that means the generated explainable paths contributed a lot to the user preference analysis and thus can be regarded as more informative. To this end,  we calculate the
Mean Squared Error (MSE) between the new score and the original score on several user-item pairs and present the average value in Table \ref{tab:l2-norm}. 

The first baseline random guess is to randomly select paths for explanations and feed them back to the recommendation backend; the second baseline attention-based module is to select explainable paths learned by the attention mechanism and also feed them to the recommendation backend. For a fair comparison, both of them have the same path numbers as our two method's average path numbers, shown by $\star$ in Table \ref{tab:l2-norm}. The last baseline, TMER \cite{chen2021temporal} is an attention-based explainable recommendation, which extracts the item-item paths as explanations of recommendations, a subset of the above two baselines' paths. Therefore, we keep the same ratio as our average explainable paths ratio for a fair comparison. The number of selected paths is shown by $\diamond$ in Table \ref{tab:l2-norm}.
Results in the table reveal that our two counterfactual-based reasoning methods contain more information for recommendations, especially the counterfactual reasoning on path structure method outperforms others. Moreover, our learned counterfactual-based explainable paths on larger datasets contribute more to the recommendation results, and other baselines also show similar trends.



\begin{table*}[ht]
\centering
\begin{threeparttable}
\resizebox{\textwidth}{!}{
{\color{black}\begin{tabular}{l|cc|cc|cc|cc}
\toprule
Datasets & \multicolumn{2}{c|}{Amazon Musical Instruments} & \multicolumn{2}{c|}{Amazon Automotive} & \multicolumn{2}{c|}{Amazon Toys and Games} & \multicolumn{2}{c}{MovieLens} \\ \midrule
\diagbox[width=10em,trim=l]{Methods}{Evaluations}                   & MSE & Num (Path) & MSE & Num (Path) & MSE & Num (Path) & MSE & Num (Path)\\ \midrule
Random Guess & 0.998  &$30^\star$  & 0.630& $18^\star$& 0.337& $11^\star$& 0.492& $16^\star$  \\ 
Attention-based Method     & 0.661      &  $30^\star$  &0.656& $18^\star$& 0.263& $11^\star$& 0.045& $16^\star$   \\ 
{TMER}     &   {0.668}    &  {$11^\diamond$}  & {\ul {0.160}} & {$6^\diamond$}& {0.044}& {$4^\diamond$}& {0.009}& {$5^\diamond$}   \\ \midrule
CR on Path Representations & \ul {0.333}   &24.0 & {\textbf{0.001}} & {21.0} & {\ul {0.056}}& {9.7}& {\ul {0.006}}& {12.0} \\ 
CR on Path Structures     & \textbf{0.261}   &37.3  & {\textbf{0.001}} & {14.3} & {\textbf{0.004}} & {13.0}& {\textbf{7e-4}}& {19.3}\\ \midrule
\bottomrule
\end{tabular}}}                                                                                                                                   
\begin{tablenotes}
        \footnotesize
        \item $\star$: Average number of our explainable paths.
        \item $\diamond$: Keep the same ratio as our explainable paths ratio.
\end{tablenotes}
\end{threeparttable}
\caption{MSE comparison between original recommendation score and only explainable paths feeding recommendation score for two perturbations on all datasets. The Num (Path) column is the number of learned explainable paths.
}\label{tab:l2-norm}
\end{table*}

\subsubsection{\textbf{Fidelity}}
\label{subsec:fedelity}

\begin{figure}[ht]
    \centering
    \includegraphics[width=0.9\linewidth]{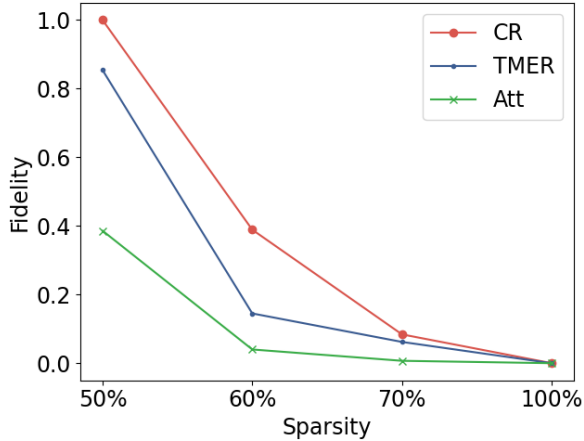}
    \caption{Fidelity of our counterfactual reasoning method and baselines. The x-axis is the sparsity of explainable paths input and the y-axis denotes fidelity scores.}
    \label{fig:fidelity} \vspace{-1em}
\end{figure}

We also conduct the fidelity experiment to compare our counterfactual reasoning with the attention module in section \ref{subsec:rec} and TMER \cite{chen2021temporal} on the Amazon Automotive dataset, following the settings in \cite{bajaj2021robust}. This metric presents the decrease in the prediction score when removing different ratios of explainable paths from the input. A larger fidelity value shows stronger counterfactual weights, indicating better explainability. The results are shown in Figure \ref{fig:fidelity}. The x-axis is the sparsity of the explainable paths, which is the ratio of the explainable paths as input. To simplify the experiment, we use the intersection of two counterfactual reasoning learned explainable paths set as the 100\% explainable paths input. It shows that our method (CR) outperforms the baselines, which proves good fidelity of our proposed counterfactual reasoning.

\subsection{Qualitative Evaluation for Explainability of our Counterfactual Reasoning ({RQ2})}
\label{subsec:exp_quali_explainability}


Recall in section \ref{subsec:eva_explain}, we propose both qualitative and quantitative ways to evaluate the explainability of different explanation methods. Here from the qualitative perspective, we compare the explainability of our proposed counterfactual reasoning with traditional attention-based explanations using our previously proposed measurements: \textit{stability}, \textit{effectiveness}, and \textit{visualization} via traditional case study evaluation. 

    

\begin{figure}[ht]
    \centering
    \includegraphics[width=0.9\linewidth]{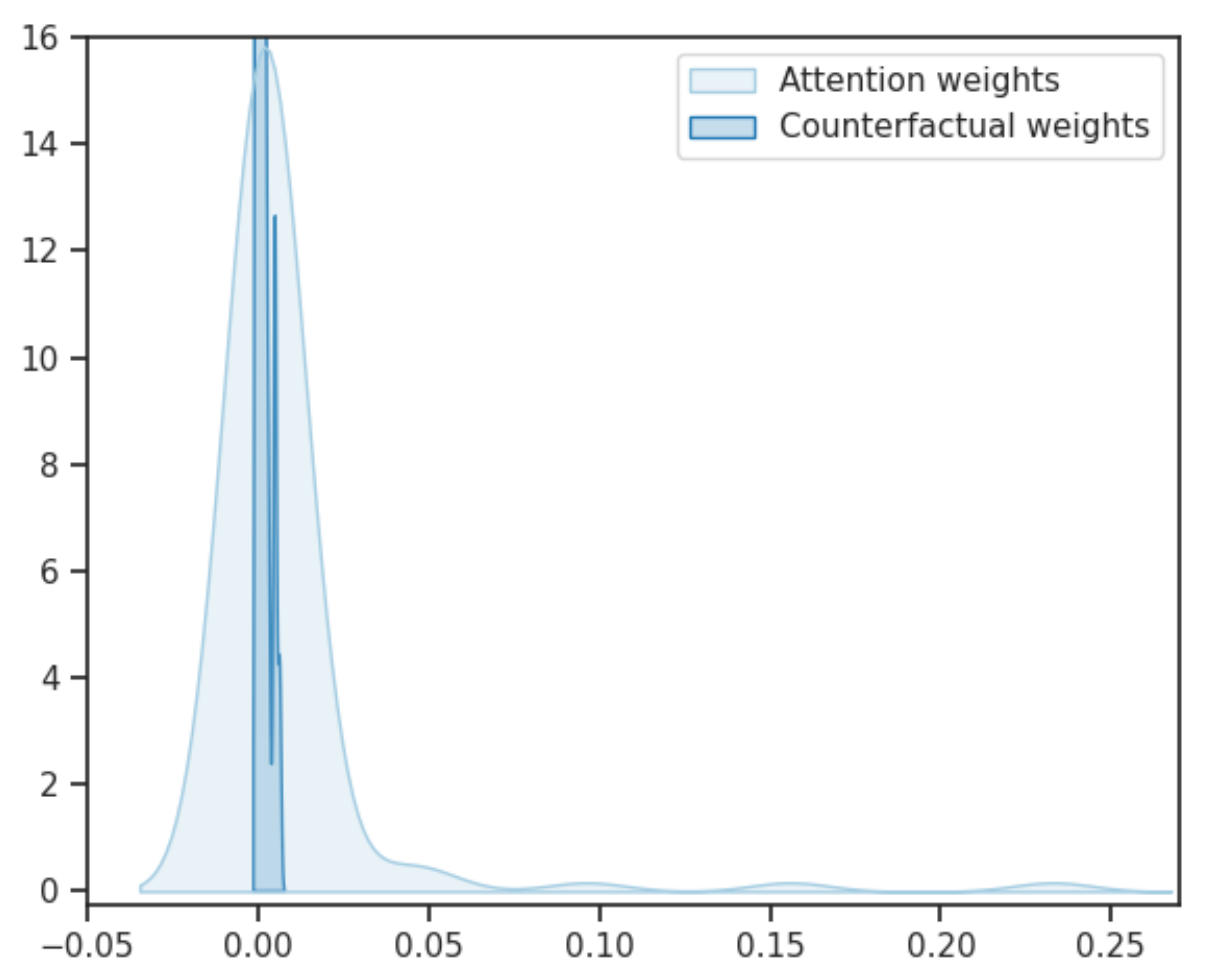}
    \caption{{Stability study. It denotes the kernel density estimation (KDE) plot of both attention-based and counterfactual-based explainable weights. }}
    \label{fig:box_figure_stability} \vspace{-1em}
\end{figure}

\subsubsection{\textbf{Stability}} \label{subsubsection:stability}
A better explanation framework should ensure their results are stable when we repeat the method multiple times because it would be ambiguous to determine the final explainable paths if the explainable model generates unstable path weights. {To this end, we randomly draw 100 explainable paths and track the learned explainable weights by our counterfactual reasoning on path representations, and compare them with attention-based weights, after 10 runnings. 
Specifically, we draw the KDE (Kernel Density Estimation) plot of the learned attention-based path weights and counterfactual path weights to demonstrate the stability. For example, the thinner KDE plot denotes lower weight variants and better stability. }
{In Figure \ref{fig:box_figure_stability}, the x-axis denotes the weights variants after ten runs. The y-axis denotes the density (frequency) of attention weights and counterfactual weights. From the figure, it is straightforward that the KDE plot of counterfactual weight is much thinner than that of the attention weights. In particular, only 16\% of paths' weights learned by the attention mechanism stay small variants, but all the counterfactual weights have small variants. This validates that our method is more stable.}

\subsubsection{\textbf{Effectiveness}} \label{subsubsection:effectiveness}

\begin{figure}[ht] 
    \centering
  \subfloat[Attention weights]{%
  \label{fig:box_figure_att_weight_10try_ridi}
       \includegraphics[width=0.8\linewidth]{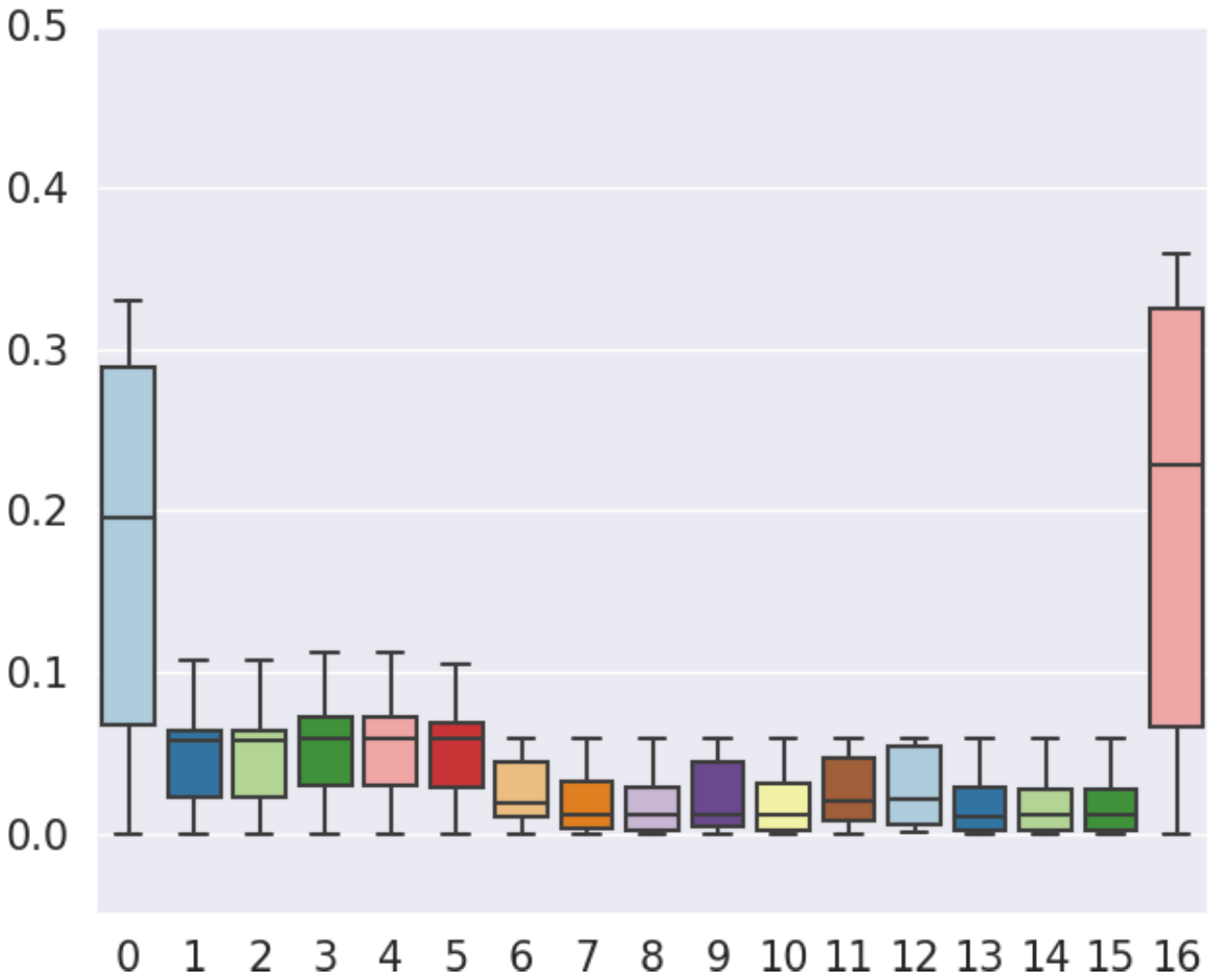}}
    \\
  \subfloat[Counterfactual weights]{%
\label{fig:box_figure_counterfactual_weight_10try_ridi}
        \includegraphics[width=0.8\linewidth]{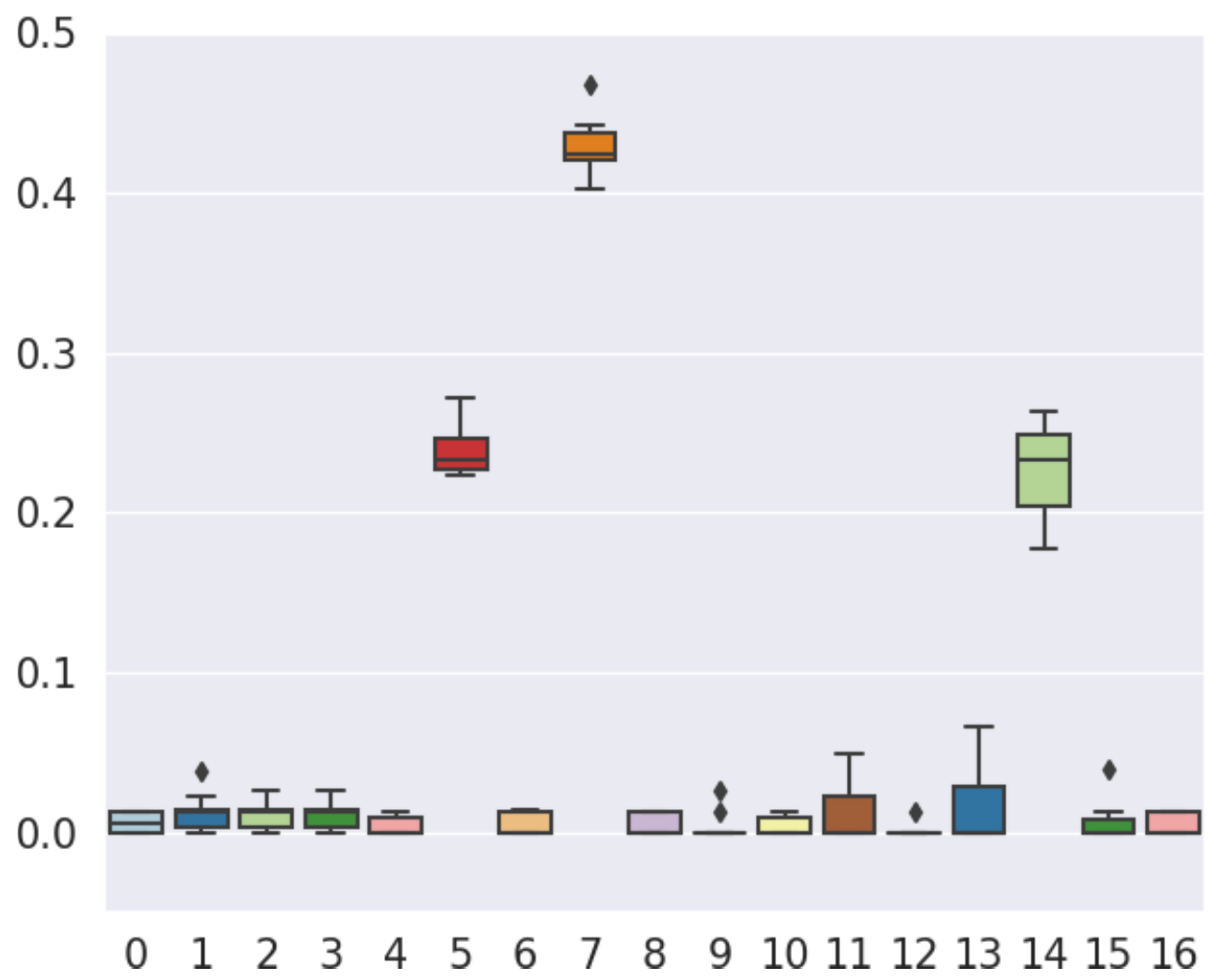}}
  \caption{Effectiveness study. The X-axis denotes the path ID and the y-axis represents the range of explainable weights learned by 10 independent runnings. (Path \#16 is the added "pranking" path.)}\vspace{-1em}
  \label{fig:box_figure_ridi} 
\end{figure}

Intuitively, if the model learns a high weight for the "pranking" path, the explanation method is less effective and less reasonable. Therefore, to simulate and see how the counterfactual reasoning model and attention mechanism learn the importance weight of "pranking" path, we randomly draw the explainable weights learned by our counterfactual reasoning model. Then we add a "pranking" path to the original path set by randomly choosing an irrelevant path (for example, a path from far distant users) to the target item and see what happens to the results.  For a fair comparison, we also independently train each explanation weight ten times. In Figure \ref{fig:box_figure_ridi}, the "pranking" paths are presented as path \#16. The x-axis denotes the path id, and the y-axis represents the learned explanation weight. The colored area of each bar indicates the range of learned weights.
From Figure \ref{fig:box_figure_att_weight_10try_ridi}, we find that the attention mechanism fails to separate the "pranking" paths from the original set. The reason may be that the irrelevant path has some common characteristics contained in the original paths. For many attention-based recommendation models, these "pranking" paths might not cause huge damage to the recommendation results, but using these paths for the explanation is far-fetched. This observation reveals the awkward gap between using an attention mechanism for recommendation performance and explanation. Compared with attention-based methods, our method is indeed better at exploring the particular path, which is informative for the explanation.

\subsubsection{\textbf{Visualization Comparison via Case Study.}}
\label{subsec:casestudy}

\begin{figure}[ht]
	\centering
        \includegraphics[width=0.44\textwidth]{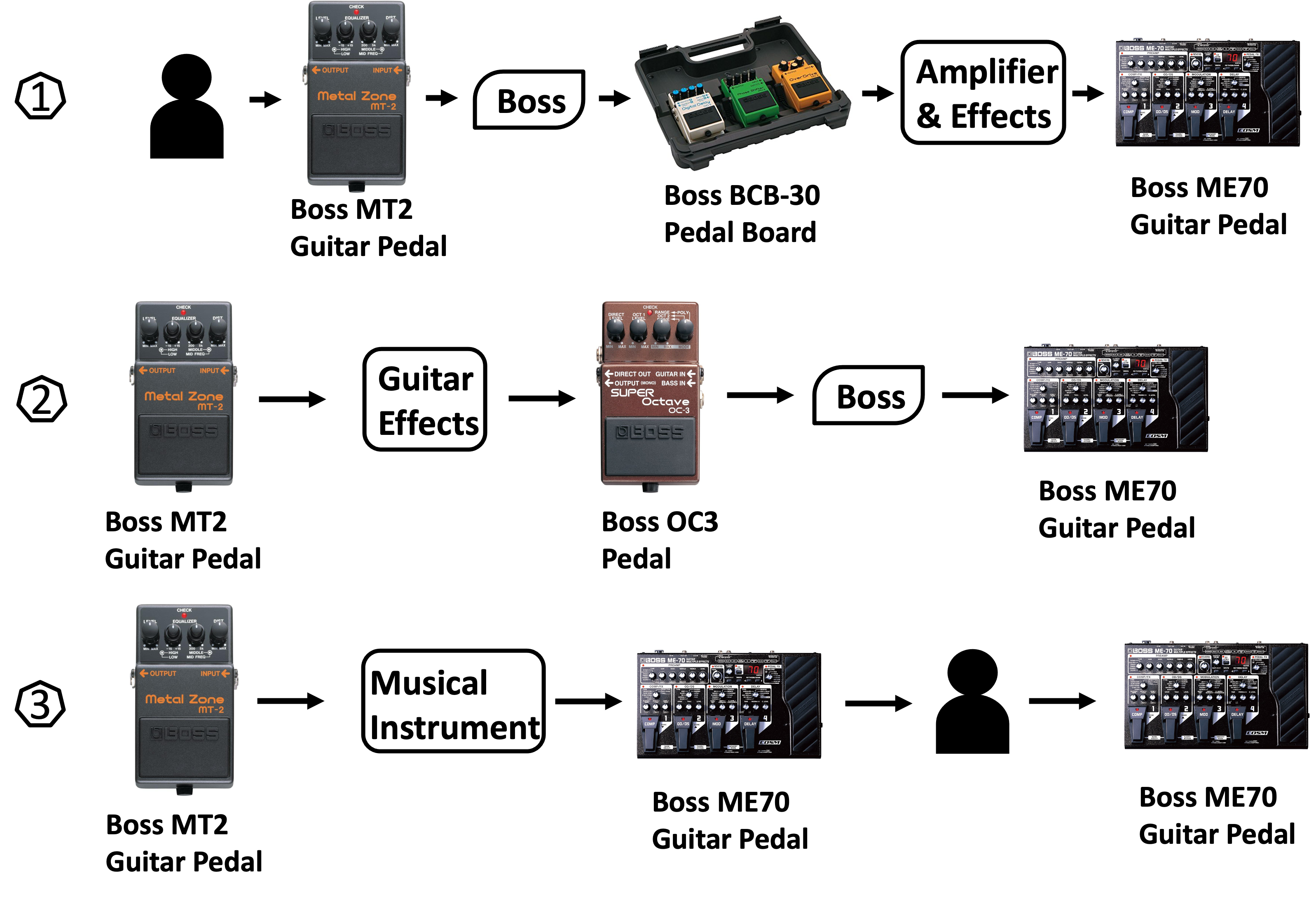}
	\caption{Case study for generating explainable paths through path representation perturbation. Boss is a brand, and others in round rectangles are the categories. Item names have been simplified to keywords because of space limitations.} \label{fig:case_study_vector_perturb}

\end{figure}

Here we randomly select two explainable path cases by our path representation perturbation and reinforcement learning-based path manipulation, respectively. In Figure \ref{fig:case_study_vector_perturb}, we randomly select a user in the Amazon Musical Instrument dataset and list her explainable paths found by our counterfactual reasoning on path representation. The normalized weights of these three paths are $[0.190, 0.189, 0.052]$, where the third path has the smallest weight. This is consistent with our commonsense because this path only contains very general information such as ``\textit{category: Musical Instrument}''. Compared with other information provided in the first two paths, ``\textit{category: Musical Instrument}'' is the widest category in the dataset. Therefore, compared with the other specific explanation path, the third path should be the worst explanation and deserves the lowest explainable weight. However, if we use attention mechanism to learn the explainable weights of these three paths, the normalized weights are $[0.0279, 0.0512, 0.0512]$, where the first path has the smallest weight.

\begin{figure}[ht]
	\centering
        \includegraphics[width=0.44\textwidth]{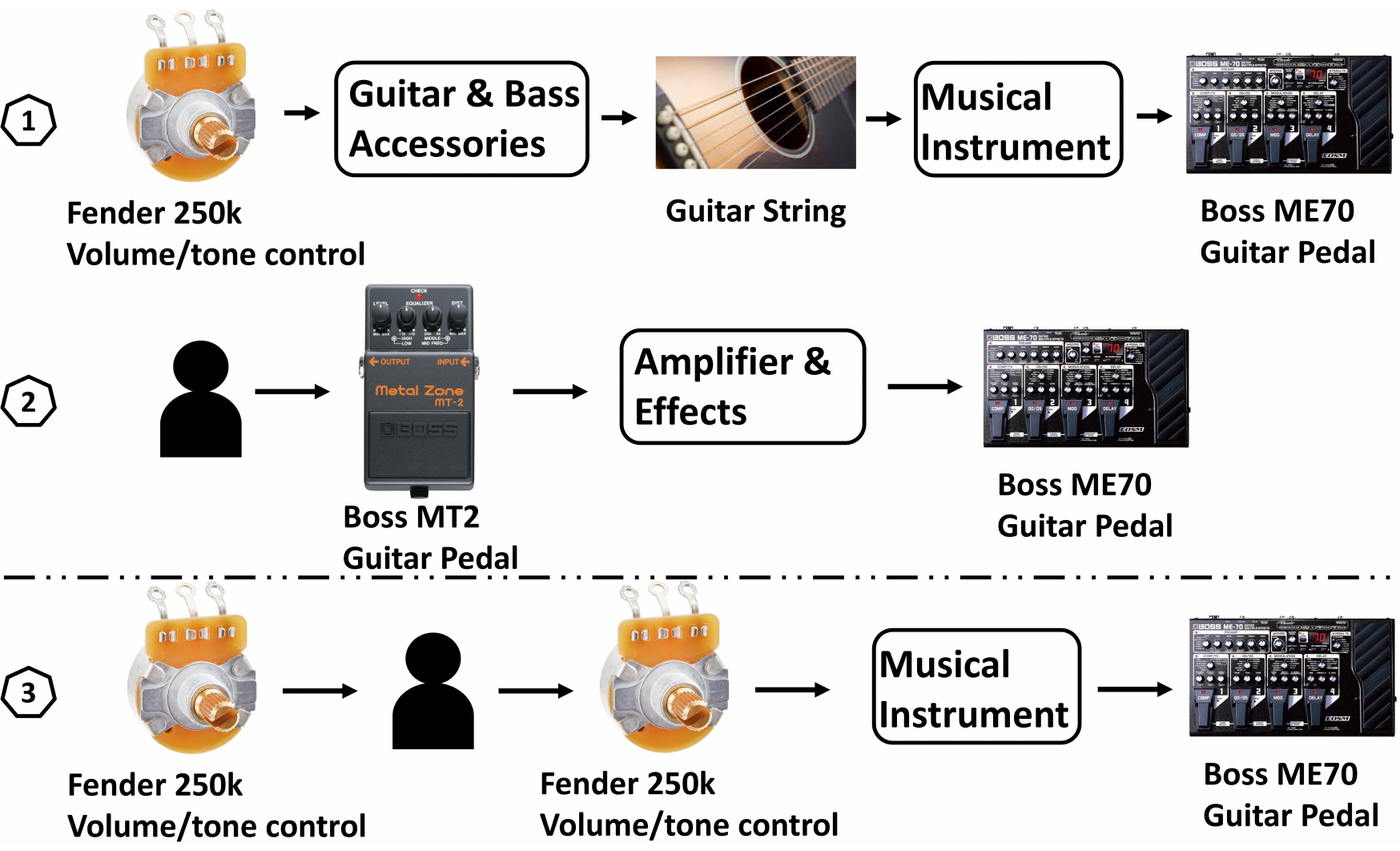}
	\caption{Case study for generating explainable paths through RL-based path manipulation. Contents in round rectangles are the categories. Item names have been simplified to the keywords because of space limitations. The first two paths are selected by RL-based counterfactual reasoning, and the last one has a high explainable weight via attention training. 
 \vspace{-2em}
	} \label{fig:case_study_rl_perturb}
\end{figure}

For the reinforcement learning-based counterfactual explainable method, we also draw the explainable paths for purchasing the ``Boss ME70 Guitar Pedal''. Note that there are some overlapping paths with our first method. To better illustrate the second method, we ignore the same paths and only present some different ones in Figure \ref{fig:case_study_rl_perturb}, where the first two paths are selected by our method but ignored by attention-based method and the third one is treated as an unimportant path by our method but with very high weight in the attention-based method. From Figure \ref{fig:case_study_rl_perturb}, we can see that the first two paths contain a very informative relation with the target item ``Boss ME70 Guitar Pedal''. The third path is more general and less informative because of the redundant item node and musical instrument category.

\subsection{Analysis on Reinforcement Learning-based Path Manipulation ({RQ3)}}
\label{subsec:CR_structure}

\begin{figure}[ht] 
    \centering
  \subfloat[Reward w.r.t. Iteration\label{fig:reward_seed3_epoch50_smooth7_linechart}]{%
       \includegraphics[width=0.7\linewidth]{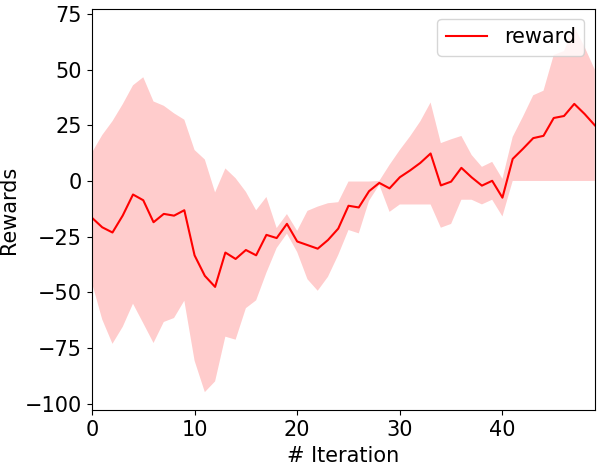}}
    \\
  \subfloat[Perturbation w.r.t. Iteration\label{fig:num_perturb_node_seed3_epoch50_smooth7_linechart}]{%
        \includegraphics[width=0.7\linewidth]{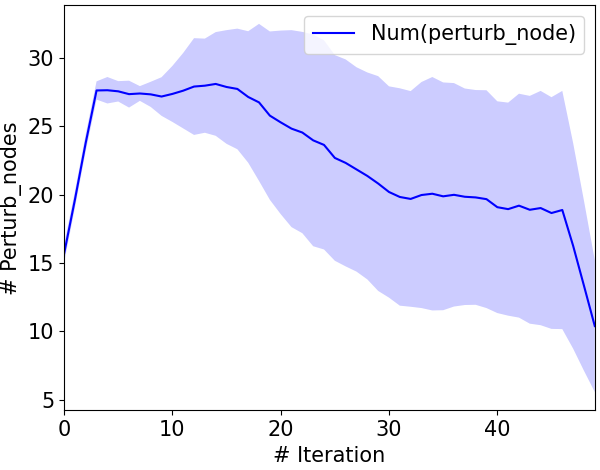}}
\caption{Reward \& Perturbation w.r.t Iteration} \label{fig:seed3_epoch50_smooth7_linechart}

\end{figure}

To present the effectiveness of the reinforcement learning model, we track the training process for a randomly selected user-item pair with 50 iterations and repeat the training process three times with different seeds. The accumulated rewards are presented in Figure \ref{fig:reward_seed3_epoch50_smooth7_linechart} where the dark color line is the mean value, and the shaded area is the range of values. Figure \ref{fig:num_perturb_node_seed3_epoch50_smooth7_linechart} presents the number of perturbed nodes. From Figure \ref{fig:reward_seed3_epoch50_smooth7_linechart}, we can see a gradual increase in the reward curve until it becomes steady. From Figure \ref{fig:num_perturb_node_seed3_epoch50_smooth7_linechart} we can see that the number of perturbed nodes drop down as the agent gradually explored the appropriate policy. These observations indicate that our proposed reinforcement learning-based agent successfully explores a path manipulation policy that tried to cause a larger result drop with a relatively small manipulation cost. To better illustrate the results from counterfactual reasoning on path representation and path structure, we further study the explainable paths generated by these two methods in the following section.

\subsection{Effectiveness Analysis on the Recommendation Model ({RQ4})}
\subsubsection{\textbf{Overall Performance}}\label{subsec:exp_overall_performance}

\begin{table*}
\centering
\resizebox{\textwidth}{!}{
{\color{black}\begin{tabular}{c|rc|ccccccccc}
\hline \toprule
Datasets                                                                             & Metrics@                       & $K$    & CountER & PGPR   & ACVAE  & Gru4Rec & BERT4Rec & SASRec & SRGNN & CPER   & Improvement \\ \hline
\multirow{8}{*}{\begin{tabular}[c]{@{}c@{}}Amazon\\ Musical \\Instruments\end{tabular}}              & \multirow{4}{*}{HR@}   & 5  & 0.0345                      & 0.0177                   & 0.0137                    & 0.0561                      & 0.0462                       & 0.0566                     & 0.0662                    & \textbf{0.0833}          & 25.83\%                         \\
                                                                                     &                        & 10 & 0.0486                      & 0.0243                   & 0.0243                    & 0.0737                      & 0.0662                       & 0.0724                     & 0.0917                    & \textbf{0.0931}          & 1.53\%                          \\
                                                                                     &                        & 20 & 0.0760                      & 0.0382                   & 0.0453                    & 0.1037                      & 0.0959                       & 0.1097                     & \textbf{0.1193}           & 0.0931                   & -21.96\%                        \\
                                                                                     &                        & 30 & 0.1254                      & 0.0512                   & 0.0674                    & 0.1254                      & 0.1159                       & \textbf{0.1386}            & 0.1379                    & 0.1098                   & -20.78\%                        \\ \cline{2-12}
                                                                                     & \multirow{4}{*}{NDCG@} & 5  & 0.0253                      & 0.0153                   & 0.0101                    & 0.0369                      & 0.0293                       & 0.0305                     & 0.0458                    & \textbf{0.0796}          & 73.80\%                         \\
                                                                                     &                        & 10 & 0.0297                      & 0.0175                   & 0.0127                    & 0.0426                      & 0.0355                       & 0.0372                     & 0.0538                    & \textbf{0.0803}          & 49.26\%                         \\
                                                                                     &                        & 20 & 0.0367                      & 0.0211                   & 0.0181                    & 0.0501                      & 0.0429                       & 0.0422                     & 0.0608                    & \textbf{0.0830}          & 36.51\%                         \\
                                                                                     &                        & 30 & 0.0434                      & 0.0239                   & 0.0228                    & 0.0548                      & 0.0472                       & 0.0466                     & 0.0648                    & \textbf{0.0864}          & 33.33\%                         \\ \hline
\multirow{8}{*}{\begin{tabular}[c]{@{}c@{}}Amazon\\ Automotive\end{tabular}}         & \multirow{4}{*}{HR@}   & 5  & 0.0227                      & 0.0131                   & 0.0099                    & 0.0586                      & 0.0504                       & 0.0435                     & 0.0670                    & \textbf{0.0797}          & 18.96\%                         \\
                                                                                     &                        & 10 & 0.0355                      & 0.0204                   & 0.0201                    & 0.0786                      & 0.0683                       & 0.0622                     & \textbf{0.0961}           & 0.0800                   & -16.75\%                        \\
                                                                                     &                        & 20 & 0.0741                      & 0.0348                   & 0.0400                    & 0.1166                      & 0.0983                       & 0.0915                     & \textbf{0.1259}           & 0.0837                   & -33.52\%                        \\
                                                                                     &                        & 30 & 0.1047                      & 0.0483                   & 0.0617                    & 0.1407                      & 0.1217                       & 0.1126                     & \textbf{0.1450}           & 0.0943                   & -34.97\%                        \\ \cline{2-12}
                                                                                     & \multirow{4}{*}{NDCG@} & 5  & 0.0147                      & 0.0111                   & 0.0070                    & 0.0413                      & 0.0336                       & 0.0306                     & 0.0469                    & \textbf{0.0797}          & 69.94\%                         \\
                                                                                     &                        & 10 & 0.0188                      & 0.0135                   & 0.0104                    & 0.0477                      & 0.0393                       & 0.0366                     & 0.0563                    & \textbf{0.0798}          & 41.74\%                         \\
                                                                                     &                        & 20 & 0.0286                      & 0.0172                   & 0.0155                    & 0.0573                      & 0.0469                       & 0.0440                     & 0.0638                    & \textbf{0.0807}          & 26.49\%                         \\
                                                                                     &                        & 30 & 0.0351                      & 0.0201                   & 0.0200                    & 0.0624                      & 0.0519                       & 0.0485                     & 0.0679                    & \textbf{0.0830}          & 22.24\%                         \\ \hline
\multirow{8}{*}{\begin{tabular}[c]{@{}c@{}}Amazon\\ Toys\\ and\\ Games\end{tabular}} & \multirow{4}{*}{HR@}   & 5  & 0.0269                      & 0.0144                   & 0.0106                    & 0.0906                      & 0.0678                       & 0.0817                     & 0.0726                    & \textbf{0.0939}          & 3.64\%                          \\
                                                                                     &                        & 10 & 0.0471                      & 0.0201                   & 0.0209                    & \textbf{0.1429}             & 0.1001                       & 0.1172                     & 0.1082                    & 0.0943                   & -34.01\%                        \\
                                                                                     &                        & 20 & 0.0715                      & 0.0315                   & 0.0411                    & \textbf{0.2006}             & 0.1455                       & 0.1528                     & 0.1539                    & 0.1005                   & -49.90\%                        \\
                                                                                     &                        & 30 & 0.1011                      & 0.0413                   & 0.0623                    & \textbf{0.2330}             & 0.1778                       & 0.1772                     & 0.1823                    & 0.1184                   & -49.18\%                        \\
                                                                                      \cline{2-12}
                                                                                     & \multirow{4}{*}{NDCG@} & 5  & 0.0155                      & 0.0128                   & 0.0075                    & 0.0585                      & 0.0437                       & 0.0537                     & 0.0475                    & \textbf{0.0939}          & 60.51\%                         \\
                                                                                     &                        & 10 & 0.0219                      & 0.0147                   & 0.0110                    & 0.0754                      & 0.0541                       & 0.0652                     & 0.0590                    & \textbf{0.0940}          & 24.67\%                         \\
                                                                                     &                        & 20 & 0.0287                      & 0.0176                   & 0.0162                    & 0.0900                      & 0.0656                       & 0.0741                     & 0.0705                    & \textbf{0.0956}          & 6.22\%                          \\
                                                                                     &                        & 30 & 0.0350                      & 0.0197                   & 0.0207                    & 0.0969                      & 0.0725                       & 0.0793                     & 0.0765                    & \textbf{0.0994}          & 2.58\%                          \\ \hline
\multirow{8}{*}{{MovieLens}}                                                           & \multirow{4}{*}{HR@}   & 5  & 0.0425                      & 0.0120                   & 0.0102                    & 0.0901                      & 0.1134                       & 0.1024                     & 0.1092                    & \textbf{0.1397}          & 23.19\%                         \\
                                                                                     &                        & 10 & 0.0782                      & 0.0123                   & 0.0200                    & 0.1633                      & 0.2001                       & 0.1738                     & 0.1986                    & \textbf{0.2486}          & 24.24\%                         \\
                                                                                     &                        & 20 & 0.1392                      & 0.0128                   & 0.0396                    & 0.2806                      & 0.3272                       & 0.2762                     & 0.3334                    & \textbf{0.4058}          & 21.72\%                         \\
                                                                                     &                        & 30 & 0.1895                      & 0.0131                   & 0.0596                    & 0.3731                      & 0.4140                       & 0.3519                     & 0.4227                    & \textbf{0.5148}          & 21.79\%                         \\ \cline{2-12}
                                                                                     & \multirow{4}{*}{NDCG@} & 5  & 0.0253                      & 0.0119                   & 0.0072                    & 0.0550                      & 0.0683                       & 0.0626                     & 0.0666                    & \textbf{0.1026}          & 50.22\%                         \\
                                                                                     &                        & 10 & 0.0368                      & 0.0120                   & 0.0106                    & 0.0785                      & 0.0961                       & 0.0856                     & 0.0953                    & \textbf{0.1397}          & 45.37\%                         \\
                                                                                     &                        & 20 & 0.0520                      & 0.0122                   & 0.0156                    & 0.1079                      & 0.1281                       & 0.1113                     & 0.1292                    & \textbf{0.1803}          & 39.55\%                         \\
                                                                                     &                        & 30 & 0.0627                      & 0.0122                   & 0.0199                    & 0.1276                      & 0.1466                       & 0.1274                     & 0.1483                    & \textbf{0.2038}          & 37.42\%                                                     \\ \bottomrule
\end{tabular}}}
\caption{{Comparison with baselines on \{HR, NDCG\}@\{5, 10, 20, 30\}. The best performance is bold, and the second-best performance is underlined. The last column presents CPER performance improvement compared with the best baseline.}}\vspace{-1em} \label{tab:baseline}
\end{table*}

{We compare our proposed recommendation model with seven state-of-the-art baselines on Hit Ratio (HR), and Normalized Discounted Cumulative Gain (NDCG). For each baseline, we sample 500 negative items and report the prediction results in Table \ref{tab:baseline}, from which we can find that our recommendation (CPER) obtains the best NDCG performance on all datasets. This means that it can hit the ground-truth label as the first ranking position with a higher probability than other comparison methods, which illustrates our recommendation can better capture the most important information.
However, it does not perform well on HR@\{10,20,30\} on three Amazon datasets, especially on Amazon Automotive and Amazon Toys and Games. The rationale is that these two datasets are sparser than Amazon Musical Instruments and MovieLens, resulting in less information on user-item interactions and paths. Meanwhile, our method is highly related to paths on the recommendation graph, which makes it more sensitive to the sparsity of the dataset. However, this shortcoming has little influence on the following module, counterfactual reasoning for explainability, because the counterfactual reasoning methods focus more on the deviation of recommendation scores.
Even if the foremost indicator of this work is not the recommendation performance, but the explainability, we admit this limitation of this path-based recommendation backend and will improve it in the future. } 


\subsubsection{\textbf{Effectiveness of Different Components}}
\label{subsec:exp_ablation_test}

\begin{table}[ht]
\centering
\begin{tabular}{@{}clll@{}}
\hline \toprule
Variants         & HR@20            & NDCG@20                        \\ \hline 
CPER             & \textbf{0.0931} & \textbf{0.0830}       \\ 
CPER$\neg$TempDW & 0.0866 ($\downarrow$ 7\%)          & 0.0797 ($\downarrow$ 4\%) \\ 
CPER$\neg$Trans  & 0.0763 ($\downarrow$ 18\%)         & 0.0440 ($\downarrow$ 47\%) \\ 
CPER$\neg$Att    & 0.0587 ($\downarrow$ 37\%)         & 0.0490 ($\downarrow$ 41\%)         & \\   \bottomrule
\end{tabular}%
\caption{Different variants result on \{HR, NDCG\}@20 for comparison. $\neg$ represents the variant without the following module. The best results are in bold.}
\label{tab:ablation_test} 
\end{table}

We further evaluate the effectiveness of the main components in our recommendation backend by comparing different variants on the Amazon Musical Instrument dataset. We report the results of \{HR, NDCG\}@20 in Table \ref{tab:ablation_test}. For the last column, we show the smallest percentage drop compared with CPER performance for each variant on two evaluation metrics.
Specifically, CPER is the complete model, which obtains the best results. CPER$\neg$TempDW removes the pre-training phrase with temporal DeepWalk, which decreases 3.98\% on NDCG@20. CPER$\neg$Trans takes the average of history item representations as the path representation instead of using a more powerful sequential model Transformer, the performance of which decreases 18.05\% on HR@20. CPER$\neg$Att means we remove all the attention layers and the related module, the item enhancement via path embeddings. 
The results then drop down by 36.95\% on HR@20. Note that most recommendation models leverage these attention units for pursuing recommendation accuracy, and therefore they indeed contribute much to the prediction results. However, they are not intentionally designed for model explainability, making the attention-based explainability very limited as discussed in the previous explainability evaluation.

\section{Related Work}

The related works are mainly divided into two aspects, including explainable recommendations and counterfactual-based reasoning.
\subsection{Explainable Recommendations}
Early explainable recommendations are mostly based on collaborative filtering (CF) relation \cite{sharma2013social}, identifying patterns and similarities between users and items based on their past interactions.
The early CF-based recommendations mostly mine the relations through matrix factorization, which can explain the user-item interaction matrix via math solution procedure. Also, this collaborative filtering relation is naturally regarded as the explanation of recommendation. 
However, these CF-based explainable recommendations have poor performance due to the deficiency in side information and composite structure modeling. They also have limited ability to provide personalized explanations since they only rely on past interactions without considering additional user or item features.
Later, thanks to the development of neural networks, the ability to learn user/item representations essentially improves, and therefore, the user/item features also could carry abundant information for explanation \cite{seo2017interpretable}. With the help of the side information, \cite{wu2015flame} also leverages the topic model for extracting explainable information. However, although the neural network-based recommendations outperform the traditional recommendations, the black-box problem is also brought by neural networks, resulting in more difficulties in explainable learning.

Nowadays, with the development of knowledge graph (KG) and graph neural networks, KG-based recommendations show a significant advantage in explainability because of the auxiliary information on graphs \cite{zhang2020explainable}. Other than rule-explanation \cite{ma2019jointly}, user/item-explanation \cite{catherine2017explainable}, user/item features-explanation \cite{huang2018improving} and sentence-explanation \cite{ai2018learning}, most recent works make full use of the graph structure and leverage underlying paths along the user-item purchase \cite{wang2019kgat, huang2021path} or item-item evolution \cite{chen2021temporal, li2023reinforcement} composite relations as an explanation, which is more informative compared with the traditional explanation. These path-based explainable recommendations always explore a large number of explainable paths. With the development and success of the attention mechanism in Natural Language Processing (NLP), many research works \cite{hu2018leveraging, li2020time} also adopt an attention mechanism to the recommendation tasks to enhance the recommendation performance. Later, it is also used to learn the importance distribution of the paths as explanations of recommendation \cite{wang2020attention, tao2021multi, chen2021temporal, li2023reinforcement}. 

However, there are some discussions about the shortcomings of attention-based explainability \cite{wiegreffe2019attention, serrano2019attention, brunner2019identifiability, grimsley2020attention} and we also summarize the points as follows. The attention mechanism is primarily designed for performance enhancement rather than explainability, which means that the quality of attention-based explainability is not guaranteed. Additionally, since the attention mechanism is closely tied to the recommendation process, different attention-based recommendations may result in varying attention weight distributions. Last but not least, the attention mechanism tends to assign higher weights to the general and less informative paths as shown in experiment \ref{subsec:casestudy}.
Therefore, to address the aforementioned issues, we propose a novel counterfactual reasoning framework that generates explainable weights.

\subsection{Counterfactual Reasoning}
Counterfactual reasoning has also come up as an alternative explainable approach \cite{verma2020counterfactual,sun2023counter, wang2021counterfactual}. It answers the 'what-if' causal question that if some condition did not occur, the result would not happen. According to the idea, the original condition is an essential reason for the result. In the early work, CountER \cite{tan2021counterfactual} proposes a model-agnostic explainable recommendation to learn slight perturbation on item aspects for counterfactual reasoning exploration. Later, CNR \cite{zhou2021intrinsic} also proposes a similar model-agnostic explainable recommendation to perform perturbation on user features for flipping the model decision to learn the counterfactual explanation. To make better use of rich data of recommendation, $\mathrm{CF^2}$\cite{xiong2021counterfactual} and CEF \cite{ge2022explainable} utilize review data of items to enrich the information for training counterfactual reasoning. 
However, the counterfactual reasoning-based recommendations mentioned above mainly focus on providing aspect or item explanations and do not take into account the use of knowledge graphs or graph neural network modeling to generate composite relations (such as paths) for the explanation.
Although PRINCE \cite{ghazimatin2020prince} proposes a path-based counterfactual explainable recommendation through removing edges as a perturbation, the explainable model is model-specific, which is inflexible for explaining other black-box recommendations. This limits the method's generalizability and may require significant modifications to be used with different models.

To deal with the aforementioned shortcomings, we propose a novel explainable framework for path-based recommendation through counterfactual reasoning. Our flexible framework can be applied to other path-based recommendation systems and generates counterfactuals by perturbing both the path representation and the path topological structure. Specifically, we introduce a novel reinforcement learning method to learn how to perturb the path topological structure through node selection and removal.

\section{Conclusion}
In this paper, we propose a path-based explainable recommendation via counterfactual reasoning, including a general path-based recommendation backend and a framework for post-hoc explanation via generating counterfactual paths. Specifically, we propose two methods to generate counterfactual paths: one is to apply slight perturbation on the path representations, and the other is to perform perturbation as little as possible on path topological structure via our designed reinforcement learning, both of whose objectives are to drop the recommendation score as much as possible. For a better comparison of the quality of explainability with traditional attention-based explanation in the recommendation, we also introduce a package of solutions for qualitative and quantitative evaluation. Through experiments, both two counterfactual explainable methods obtain more reliable and trustworthy explainable paths compared with the traditional explainable attention weights.

In our future work, there are three directions to research for better explainability of the recommendation. The first one is to explore the question of how far will the counterfactual reasoning go in the recommendation scenario or in graph neural networks. Another is to explore more mechanisms other than attention and counterfactual reasoning for better explainability. The last important direction is to design a more general matrix-based evaluation for explainability. In a word, there is still a long way to go in the explainability of recommendations or even other scenarios.

\section*{Acknowledgments}
This work was partially conducted in the Research Institute for Artificial Intelligence of Things (RIAIoT) at PolyU. It is supported by the Australian Research Council (ARC) under Grant No. DP200101374, LP170100891, DP220103717, LE220100078, and PolyU Research and Innovation Office under Grant No. BD4A.

%
\bibliographystyle{IEEEtran}
\begin{IEEEbiography}[{\includegraphics[width=1in,height=1.25in,clip,keepaspectratio]{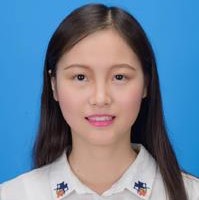}}]{Yicong Li}
is currently a Ph.D. student of the Data Science and Machine Intelligence (DSMI) Lab of Advanced Analytics Institute, University of Technology Sydney. Her research interests mainly focus on data science, graph neural networks, recommender systems, natural language processing and so on. In particular, her current research is focusing on explainable machine learning, especially the application in the recommendation area. She has published papers in international conferences and journals, such as TKDE, WSDM, CIKM, KSEM and IEEE Access. She has also reviewed papers in many top-tier conferences and journals, like AAAI, KDD, WWW, IJCAI, WSDM, ICONIP and so on. In addition, she has been invited to review manuscripts in IEEE Transactions on Neural Networks and Learning Systems (TNNLS), which is a top-tier journal in artificial intelligence. 
\end{IEEEbiography}

\begin{IEEEbiography}[{\includegraphics[width=1in,height=1.25in,clip,keepaspectratio]{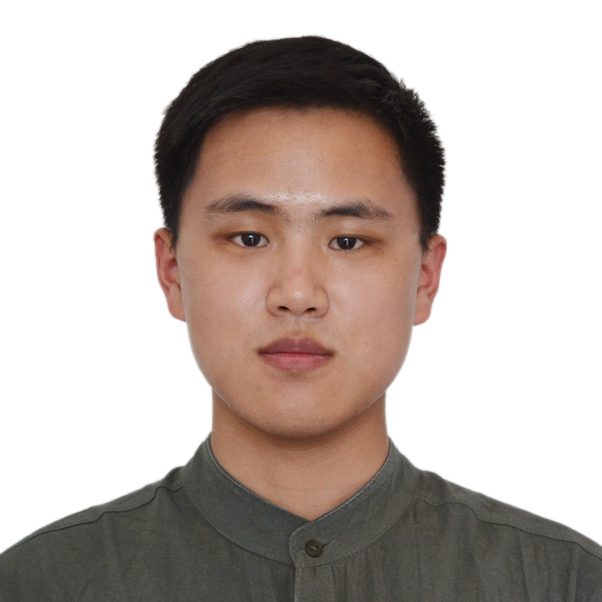}}]{Xiangguo Sun}
is a postdoctoral research fellow at The Chinese University of Hong Kong. He was recognized as the "Social Computing Rising Star" in 2023 by CAAI. He studied at  Zhejiang Lab as a visiting researcher in 2022. In the same year, he received his Ph.D. from Southeast University and won the Distinguished Ph.D. Dissertation Award. During his Ph.D. study, he worked as a research intern at Microsoft Research Asia (MSRA) from Sep 2021 to Jan 2022 and won the ''Award of Excellence''. He studied as a joint Ph.D. student at The University of Queensland hosted by ARC Future Fellow Prof. Hongzhi Yin from Sep 2019 to Sep 2021. His research interests include social computing and network learning. He was the winner of the Best Research Paper Award at KDD'23, which is the first time for Mainland and Hong Kong of China. 
He has published 12 CORE A*, 10 CCF A, and 15 SCI (including 8 IEEE Trans), some of which appear in SIGKDD, VLDB, The Web Conference (WWW), TKDE, TOIS, WSDM, TNNLS, CIKM, etc.
\end{IEEEbiography}
\begin{IEEEbiography}[{\includegraphics[width=1in,height=1.25in,clip,keepaspectratio]{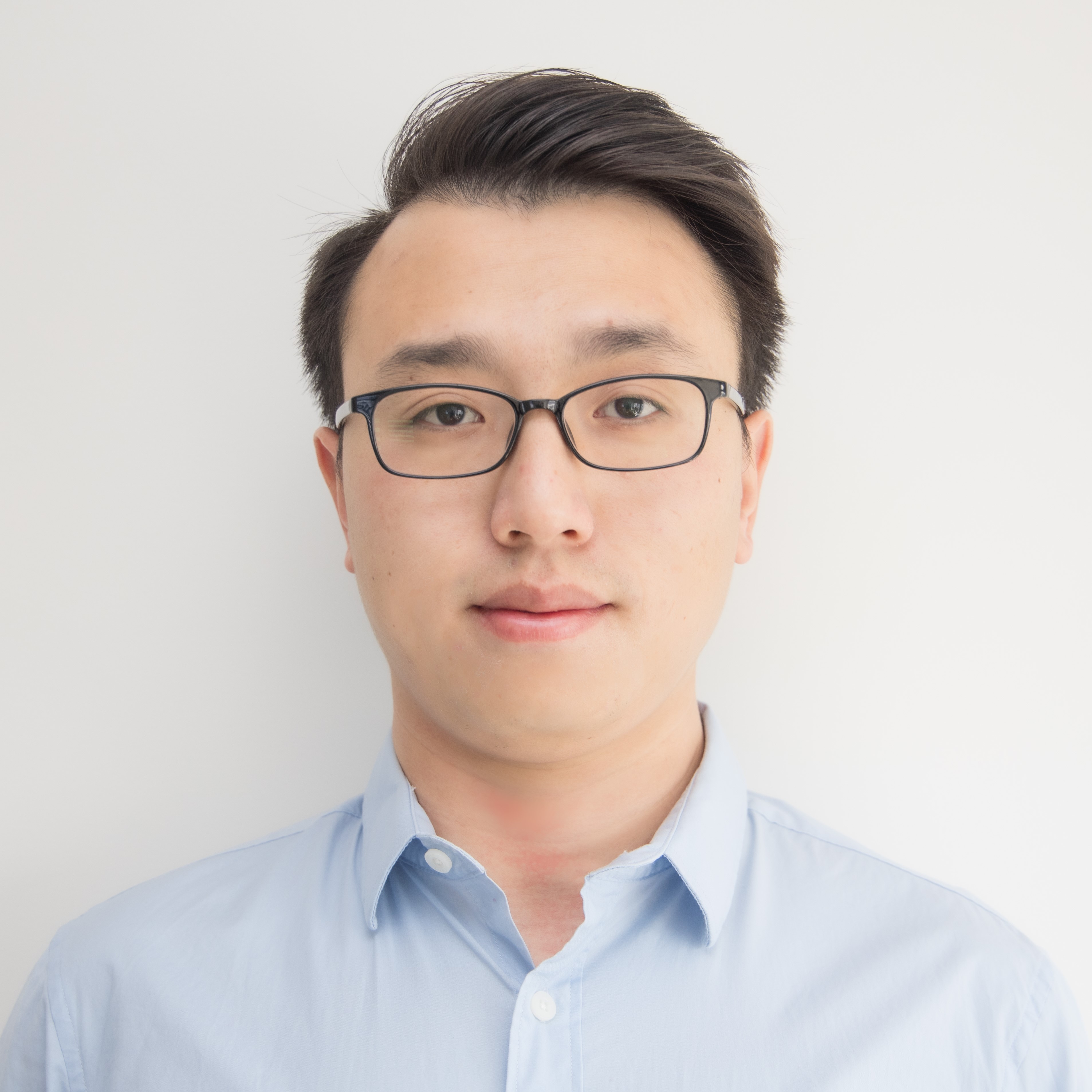}}]{Hongxu Chen}
is a Data Scientist, now working as a Postdoctoral Research Fellow in School of Computer Science at University of Technology Sydney, Australia. He obtained his Ph.D. in Computer Science at The University of Queensland in 2020. His research interests mainly focus on data mining, graph representations, social network analytics and so on. In particular, his research is focusing on learning representations for information networks and applying the learned network representations to solve real-world problems in complex networks such as biology, e-commerce and social networks, financial market and recommendations systems with heterogeneous information sources. He has published many papers in top-tier conferences and journals, such as SIGKDD, ICDE, AAAI, IJCAI, TKDE and so on. He also serves as program committee member and reviewers in multiple international conference, such as KDD, SIGIR, AAAI, and so on, and he also acts as invited reviewer for multiple journals in his research fields, including Transactions on Knowledge and Data Engineering (TKDE), VLDB Journal, IEEE Transactions on Systems and so on.
\end{IEEEbiography}
\begin{IEEEbiography}[{\includegraphics[width=1in,height=1.25in,clip,keepaspectratio]{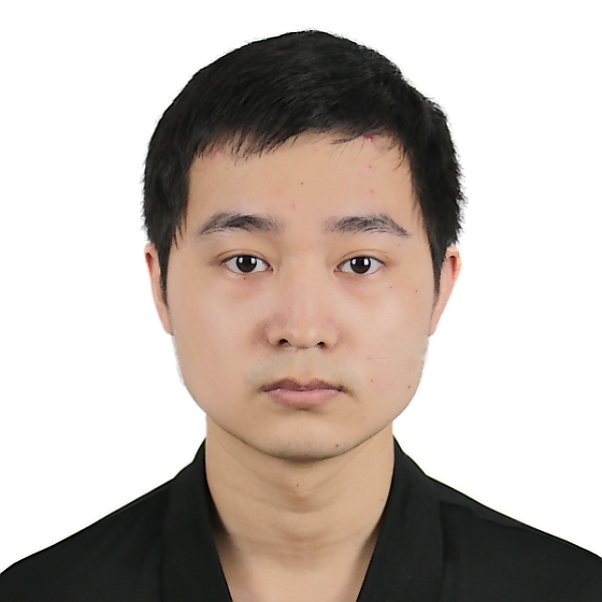}}]{Sixiao Zhang}
is currently a PhD student at Nanyang Technological University. He obtained his master's degree from Case Western Reserve University in the US, and obtained his bachelor's degree from University of Science and Technology of China. He is interested in the broad graph mining area and its downstream applications, including graph neural networks, recommender systems, adversarial robustness, etc. He has published papers in top-tier conferences and journals including KDD, WSDM, WWW, etc.
\end{IEEEbiography}
\begin{IEEEbiography}[{\includegraphics[width=1in,height=1in,clip,keepaspectratio]{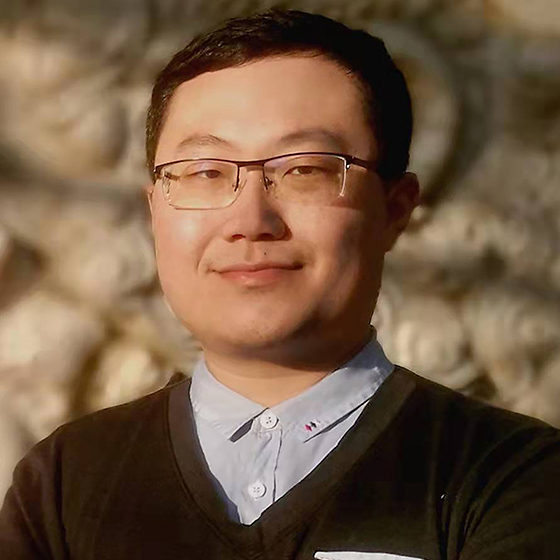}}]{Yu Yang}
is currently a Research Assistant Professor with the Department of Computing, The Hong Kong Polytechnic University. He received the M.Eng. degree in Pattern Recognition and Intelligence System from Shenzhen University in 2015 and the Ph.D. degree in Computer Science from The Hong Kong Polytechnic University in 2021. His research interests include spatiotemporal data analysis, representation learning on dynamic graphs, urban computing, and learning analytics.
\end{IEEEbiography}
\begin{IEEEbiography}[{\includegraphics[width=1in,height=1.25in,clip,keepaspectratio]{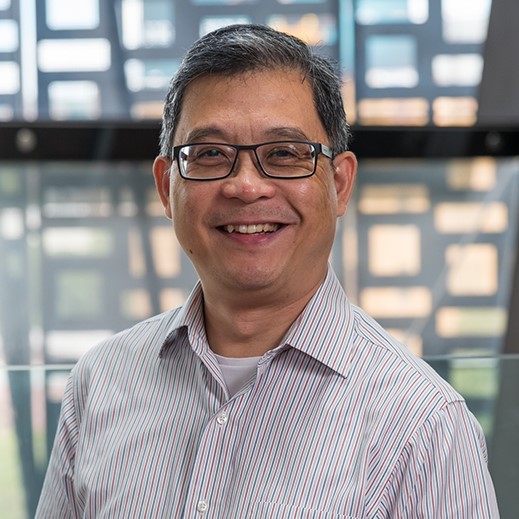}}]{Guandong Xu}
is a professor in the School of Computer Science and Data Science Institute at UTS and an award-winning researcher working in the fields of data mining, machine learning, social computing and other associated fields.
He is Director of the UTS-Providence Smart Future Research Centre, which targets research and innovation in disruptive technology to drive sustainability. 
He also heads the Data Science and Machine Intelligence Lab, which is dedicated to research excellence and industry innovation across academia and industry, aligning with the UTS research priority areas in data science and artificial intelligence.
Guandong has had more than 220 papers published in the fields of Data Science and Data Analytics, Recommender Systems, Text Mining, Predictive Analytics, User behavior modeling, and Social Computing in international journals and conference proceedings in recent years, with increasing citations from academia. 
\end{IEEEbiography}

\end{document}